\documentclass[apajournal,article,submit,pdftex,moreauthors]{mdpi} 

\firstpage{1} 
\pubvolume{1}
\issuenum{1}
\articlenumber{0}
\pubyear{2025}
\copyrightyear{2025}
\datereceived{ } 
\daterevised{ } 
\dateaccepted{ } 
\datepublished{ } 
\hreflink{https://doi.org/} 

\usepackage{xcolor}
\usepackage{bm}
\usepackage{tabularx}
\usepackage{verbatim}

\usepackage{natbib}

\Title{Long-Term Probabilistic Forecast of Vegetation Conditions Using Climate Attributes in the Four Corners Region}

\TitleCitation{Long-Term Probabilistic Forecast of Vegetation Conditions by Climate Attributes in the Four Corners Region}

\Author{Erika McPhillips $^{1}$, Hyeongseong Lee $^{1}$, Xiangyu Xie $^{1}$, Kathy Baylis $^{2}$, Chris Funk $^{3}$ and Mengyang Gu $^{1,}$*}

\AuthorNames{Erika McPhillips, Hyeongseong Lee, Xiangyu Xie, Kathy Baylis, Chris Funk, and Mengyang Gu}

\AuthorCitation{McPhillips, E., Lee, H., Xie, X., Baylis, K., Funk, C., \& Gu, M.}

\address{%
$^{1}$ \quad Department of Statistics and Applied Probability, University of California Santa Barbara, CA, USA\\
$^{2}$ \quad Department of Geography, University of California Santa Barbara,  CA, USA\\
$^{3}$ \quad Climate Hazards Center, University of California Santa Barbara, CA, USA}

\abstract{Weather conditions can drastically alter the state of crops and rangelands, and in turn, impact the incomes and food security of individuals worldwide. Satellite-based remote sensing offers an effective way to monitor vegetation and climate variables on regional and global scales. The annual peak Normalized Difference Vegetation Index (NDVI), derived from satellite observations, is closely associated with crop development, rangeland biomass, and vegetation growth. Although various machine learning methods have been developed to forecast NDVI over short time ranges, such as one-month-ahead predictions, long-term forecasting approaches, such as one-year-ahead predictions of vegetation conditions, are not yet available. 
To fill this gap, we develop a two-phase machine learning model to forecast the one-year-ahead peak NDVI over high-resolution grids, using the Four Corners region of the Southwestern United States as a testbed. In phase one, we identify informative climate attributes, including precipitation and maximum vapor pressure deficit, and develop the generalized parallel Gaussian process that captures the relationship between climate attributes and NDVI. In phase two, we forecast these climate attributes using historical data at least one year before the NDVI prediction month, which then serve as inputs to forecast the peak NDVI at each spatial grid. We developed open-source tools that outperform alternative methods for both gross NDVI and grid-based NDVI one-year forecasts, providing information that can help farmers and ranchers make actionable plans a year in advance.  }


\addhighlights{yes}
\renewcommand{\addhighlights}{%
\nolinenumbers

\noindent\textbf{What are the main findings?}
\begin{itemize}[labelsep=2.5mm,topsep=-3pt]
\item Develop a two-stage framework that integrates remotely sensed climate attributes and vegetation indices to obtain one-year-ahead probabilistic forecasts of fine-grained vegetation conditions. 

\item The climate attributes, such as precipitation and vapor pressure deficit, are found to have distinct temporal correlation patterns, which can be captured by the proposed model to improve long-term forecast accuracy. 

\end{itemize}\vspace{3pt}
\textbf{What is the implication of the main finding?}
\begin{itemize}[labelsep=2.5mm,topsep=-3pt]
\item 
Fine-grained, one-year-ahead forecasts of vegetation conditions require flexible models that can estimate the weak temporal correlations of climate attributes and vegetation indices across consecutive years over a large number of spatial grids.

\item Allows farmers and ranchers to anticipate vegetation and climate conditions up to a year in advance with quantified uncertainty for strengthened food and economic security.
\end{itemize}

}

\begin{document}
\nolinenumbers


\section{Introduction}
Changes in weather trends and extreme events have substantial impacts on food and vegetation production worldwide.
For instance, the Four Corners region of the United States (U.S.), an area covering parts of Arizona, Utah, Colorado, and New Mexico, has experienced severe droughts in recent years \citep{williams2023anthropogenic},
resulting in significant loss of vegetation and crops \citep{dannenberg2022exceptional}. The unpredictable loss of vegetation places tremendous strain on cattle and sheep ranchers, whose herds depend on shrubs and grasses for grazing \citep{redsteer2013unique}. Ranching is a significant source of income in this region, particularly for the Hopi Tribe and Navajo Nation \citep{ferguson2010drought, redsteer2013unique}. 

Satellite-based remote sensing provides an effective way to monitor vegetation conditions at high spatial resolutions over time. The Normalized Difference Vegetation Index (NDVI), a remote-sensed indicator of vegetative greenness \citep{tucker1979red}, is used to quantify vegetation conditions, assess crop yield, and rangeland biomass. Annual peak values of NDVI are strongly correlated with the aboveground net primary production (ANPP) in the U.S. Great Plains \citep{chen2019assessing}, and thus forecasting the peak of NDVI becomes a crucial step in predicting crop and vegetation production amount \citep{hartman2020seasonal}. Since NDVI is sensitive to climate attributes such as precipitation and vapor pressure deficit (VPD), the difference between the water vapor pressure at saturation (100 percent relative humidity) and the actual water vapor pressure at a given air temperature \citep{yuan2019increased}, integrating climate information into NDVI forecasting is essential. 

A range of machine learning approaches have been developed for short-term forecasting of vegetation conditions, typically for one-month-ahead lead times ~\citep{fathollahi2024global, xu2024monthly, roy2021optimum}. 
Deep learning methods, including 
convolutional neural networks \citep{lecun2002gradient} with time series decomposition \citep{gao2023ndvi},  
bidirectional long short-term memory network (LSTM)
\citep{farbo2024forecasting}, and transformer-based multi-modal deep learning framework
\citep{benson2024multi},  make use of large training datasets and strong temporal correlation between two consecutive time points. Some of these approaches extend forecasting capabilities to three months in advance. 
 Other than deep learning approaches, \cite{zhou2025enhanced} enhanced the accuracy of monthly NDVI predictions by employing a linear combination of outputs from individually optimized statistical and machine learning models, namely linear regression, support vector machines (SVM), k-nearest neighbors (KNN), random forest, and XGBoost. Overall, while these methods provide reliable short-term forecasts, they do not address longer-term forecasting (e.g., one year ahead), and the uncertainty of the prediction is not always made available.

For long-term forecasts, Grass-Cast \citep{hartman2020seasonal} provides county-level forecasts of NDVI using linear models, and in turn county-level aboveground net primary production (ANPP) under three precipitation scenarios. However, it depends on the user's hypothetical projections of future weather conditions. Therefore, reliable long-term NDVI forecasts that incorporate predicted climate attributes, with flexible models and uncertainty quantification, can provide probabilistic and actionable information.

Peak NDVI exhibits almost zero  correlation between two consecutive years, which limits the effectiveness of machine learning (ML) approaches that rely only on NDVI from the previous year. To overcome this limitation, it is crucial to identify  climate variables that are more predictable and more strongly correlated with NDVI. 
VPD, for example, has been shown to be negatively associated with the NDVI patterns globally \citep{yuan2019increased}, and we found that VPD can be predicted more accurately from data in prior years. 
Precipitation is another crucial factor influencing peak NDVI. The negative impacts of drought on NDVI have been studied in \cite{el2018vegetation,el2018characterizing} using linear models. 
Furthermore, using linear models, \cite{williams2023anthropogenic} identified July-August average VPD and January-August average precipitation to have a strong impact on NDVI in the Four Corners region. 

The goal of this paper is to forecast peak NDVI one year ahead and to quantify the uncertainty associated with the forecast, thereby enabling early planning for farmers and ranchers. Long-term forecasting of NDVI is a particularly challenging task due to 
the high stochasticity of NDVI across different spatial grids and the weak correlation between two consecutive years. Our contributions are threefold. First, we develop a generalized parallel partial Gaussian process (G-PPGP)  model, which is flexible and computationally efficient, to capture the association between climate variables and peak NDVI with quantified uncertainty. We also incorporate variable selection methods that identify precipitation and VPD in specific time ranges as the variables most correlated with peak NDVI, making them informative predictors. 
Second, we found VPD has a positive temporal correlation, while precipitation shows a negative temporal correlation pattern for the Four Corners region, which may be attributed to the extreme weather conditions in recent years. 
To accommodate these differences, we extend our model 
to allow for distinct correlation structures for precipitation and VPD, improving the accuracy and flexibility of one-year-ahead climate variable forecasts. Third,  
we forecast these climate attributes in the selected time ranges and use the forecasts as inputs to the G-PPGP model to predict NDVI. As a result, this approach produces better performing one-year-ahead forecasts that can detect NDVI fluctuations that would typically go unnoticed in other models relying solely on historical (non-predicted) covariates. Early, reliable forecasts can help farmers and ranchers make plans and take action earlier in the year. Lastly, our method is applicable to other regions and remote sensing datasets.  


The article is organized as follows. In Section \ref{subsec:data}, we introduce the region and data in this study. Section \ref{subsec:strategy} discusses the general two-phase strategy to predict NDVI and the general conceptual framework for long-term NDVI forecasting. Next, Section \ref{subsec:feature_selection} discusses the two methods of feature selection and their result. In Section \ref{subsec:attribution}, we introduce generalized parallel partial Gaussian stochastic process, linear regression, and various neural network methods for attribution modeling.
Section \ref{subsec:one-year-ahead_cov} discusses the correlations of climate attributes and their forecasting methods. 
 Section \ref{ndvi_forecast} discusses approaches for long-term NDVI forecasting. 
In Section \ref{results}, we present the prediction results from the attribution models and the one-year-ahead forecast of gross NDVI and the values at each spatial grid.  Section \ref{discussion} compares  the performance of each model and provides insight about the results.
We conclude the paper with Section \ref{conclusion}, where we discuss possible future extensions to our work. The data and code used in this work are made publicly available (\url{https://github.com/UncertaintyQuantification/forecast_ndvi}).

\section{Materials and Methods}

\subsection{Study Region and Data}
\label{subsec:data}
\begin{figure}[H]
    \centering
    \includegraphics[width=.9\textwidth]{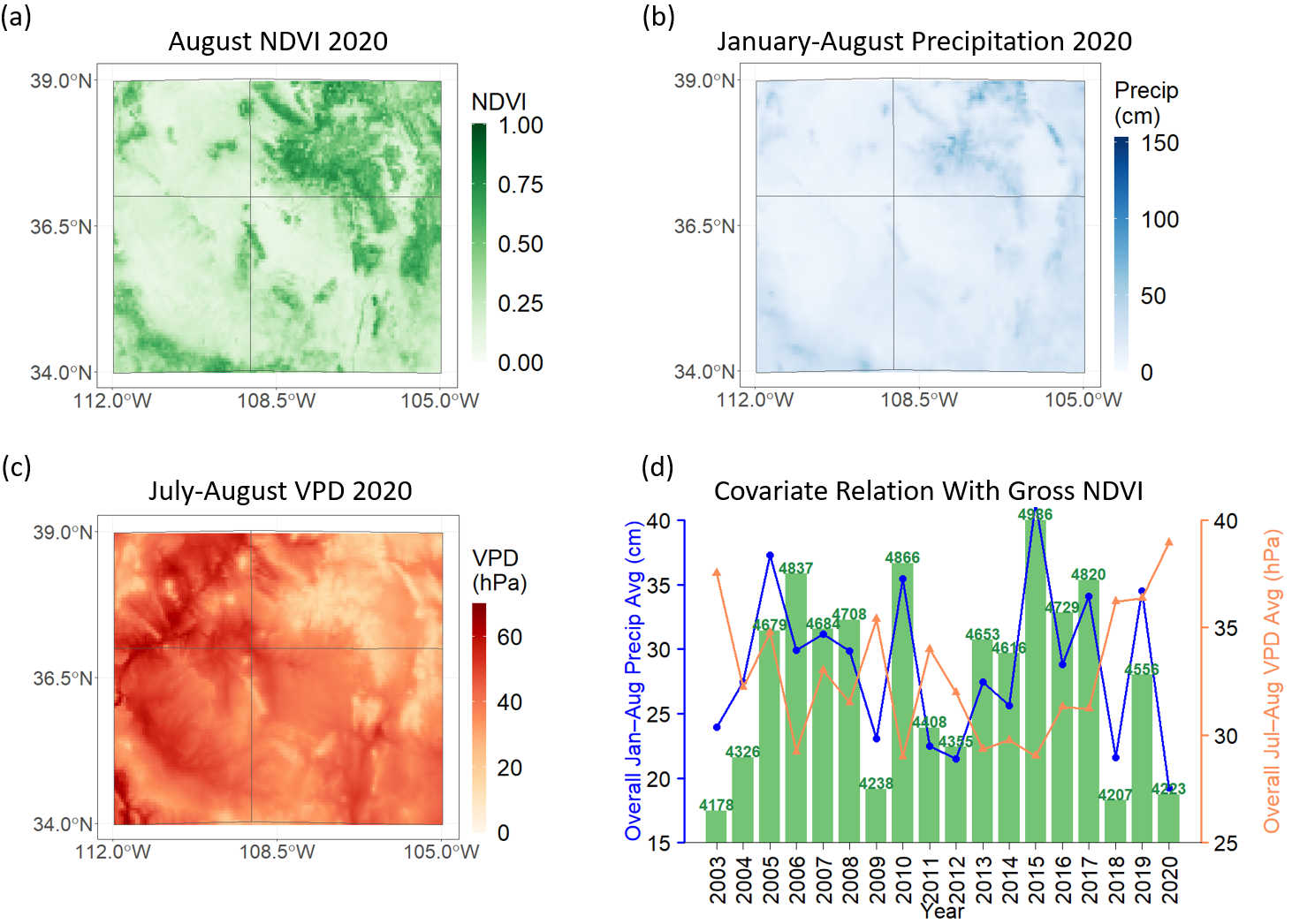}
    \caption{(a) Map of August NDVI values for the year 2020 in the Four Corners region. The longitude and latitudes are displayed on the edges of the plots. The gray horizontal and vertical lines within the plots are the inner borders of Utah, Colorado, Arizona, and New Mexico. (b) January-August average precipitation for the year 2020. (c) July-August average Max VPD for the year 2020. (d) The dark blue line represents the average January-August precipitation over the whole region, with the respective axis on the left y-axis. The green histogram is the gross NDVI (NDVI summed over all locations). The orange line is the average July-August max VPD over the whole region, with the respective axis on the right y-axis.}
    \label{fig:data_image}
\end{figure}

We use the Four Corners region of the United States (U.S.) as a testbed, which encompasses latitudes 34 to 39 degrees North and longitudes 112 to 105 degrees West. This region covers the intersection of Colorado, Utah, Arizona, and New Mexico. Using remote sensing data enables spatio-temporal assessment of vegetation and climate variables across the region. The Normalized Difference Vegetation Index (NDVI) data, with a resolution of 0.05 degrees latitude by 0.05 degrees longitude, were sourced from the Moderate Resolution Imaging Spectroradiometer (MODIS)~\citep{didan2015multi} from the United States Geological Survey (USGS) Earth Data platform. Additionally, we utilized monthly precipitation, measured in centimeters, and daily
maximum vapor pressure deficit (VPD) averaged over all days of the month, measured in hPa, as the two main climate attributes to forecast the peak NDVI in August. These climate variables, derived from remotely sensed data, were acquired from the PRISM Climate Group at Oregon State University \citep{di2008constructing} at a spatial resolution of 0.04 degrees latitude by 0.04 degrees longitude. All datasets were reprojected to the WGS84 coordinate system and resampled to match the NDVI dataset's resolution with bilinear interpolation. The resulting dataset is composed of 216 monthly data points from 2003 to 2020, where each month forms a map of 14,000 cells. Each of these cells is approximately 0.05 degrees in latitude and longitude, which is about $5.6 \text{km} \times 4.5 \text{km}$. Examples of this data are shown in Figure \ref{fig:data_image}(a)-(c), where averages of NDVI, precipitation, and VPD are shown in the Four Corners region for the year 2020.

\unskip

\subsection{A Two-Stage Framework for Long-Term NDVI Forecasting}
\label{subsec:strategy}

Although a wide range of statistical and machine learning approaches have been used to predict vegetation in the short-term \citep{lee2024contrasting, cui2020forecasting}, farmers and ranchers often require long-term predictions to prepare for the coming year. In this study, we aim to forecast the August (peak) NDVI using remotely sensed data one year ahead at a fine spatial scale (0.05 degrees), which is much more challenging than forecasting NDVI in the short-term. 

Let $\mathbf y(t_j)=(y_1(t_j),\ldots,y_k(t_j))^T$ denote the vector of August NDVI in year $t_j$ over $k$ spatial grids, where $k=100\times 140$, covering the Four Corners region. A conventional way of forecasting is to learn a one-step-ahead transition function where the inputs are 
 the August NDVI values starting from the first available year up to year $t_{j-1}$, other remotely sensed or gridded climate attributes such as precipitation and VPD also from the first year up to year $t_{j-1}$, and a noise vector at time $t_j$. Several strategies, including linear state-space models like vector autoregressive models \citep{petris2009dynamic,prado2010time} and nonlinear machine learning approaches including long short-term memory (LSTM) architecture  \citep{hochreiter1997long} and Gaussian processes \citep{gu2024probabilistic}, perform well for short-term forecasting when the output data at  two consecutive time points are strongly correlated. However, these methods typically do not perform well for long-term forecasts of peak NDVI because their temporal correlation across consecutive years is around zero.

\begin{figure}[t]
    \centering
    \includegraphics[width=1\textwidth]{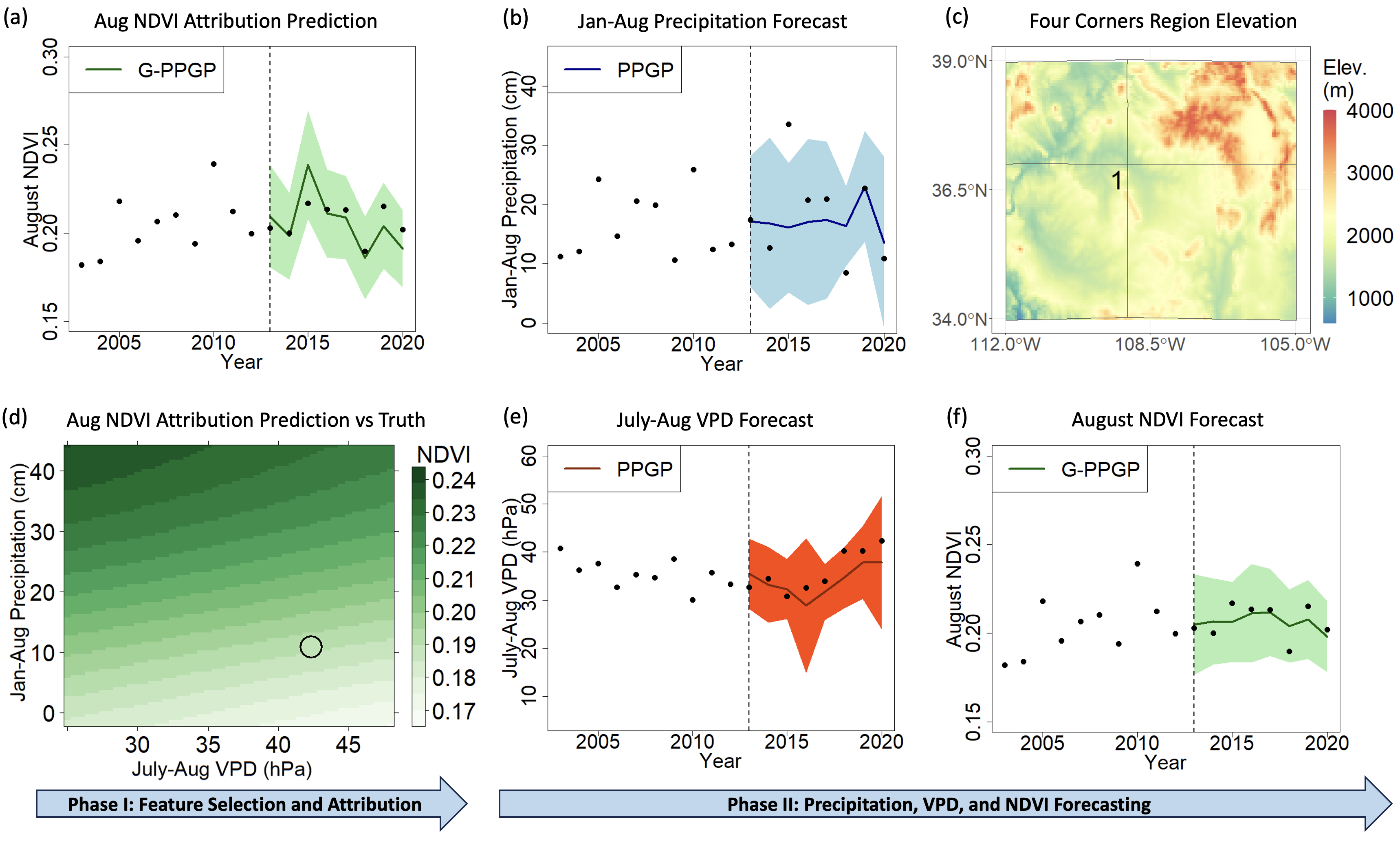}
    \caption{Illustration of the two-phase model to obtain one-year-ahead forecasts of August NDVI.
    (a) The prediction by the attribution model where August NDVI is predicted using G-PPGP assuming precipitation and VPD are known. The black dots are the true observed August NDVI at location "1" and the dark green line is the model prediction. The light green band is the 95\% credible interval (CI). (b) The forecast model for January-August average precipitation one year ahead at location "1". The dark blue line is the PPGP prediction, and the light blue band is the 95\% CI. (c) Elevation heatmap of the Four Corners region. "1" marks the sample location in the region. (d) Heatmap of the attribution model prediction at "1" given various precipitation and VPD inputs. The location of the circle indicates the true climate attribute values during 2020, and the color of the circle indicates the true August NDVI value observed during 2020.  (e) The forecast for the July-August average VPD one year ahead at location "1". The dark orange line is the PPGP prediction, and the orange band is the 95\% CI. (f) The forecast of August NDVI one-year-ahead at location "1", where the dark green line is the G-PPGP prediction and the light green band is the 95\% CI.
    }
    \label{fig:flowchart}  
\end{figure}

To address the weak temporal correlation of annual NDVI, we design a two-phase modeling framework illustrated in Figure \ref{fig:flowchart}. In the first phase, we determine which months of precipitation and VPD, prior to or during August, are most informative for predicting August NDVI, and build an attribution model using these covariates to predict August NDVI. This initial phase demonstrates that precipitation and VPD are informative predictors 
and establishes the necessary inputs and model structure for the next phase. In the second phase, we forecast the precipitation and VPD covariates one year in advance using data up to August from previous years, which then serve as inputs to a predictive model to create one-year-ahead forecasts of peak NDVI across all spatial grids. 
This strategy takes advantage of the relatively strong temporal correlation of the climate variables, as well as their association with peak NDVI, which helps overcome the limitation that peak NDVI has a weak year-to-year correlation.

 

\subsection{Climate Variable Selection and Attribution Model}\label{subsec:attribution}

The goal of the first phase is to create an accurate attribution model, a predictive model of NDVI based on climate variables from the same year, before extending it to one-year-ahead forecasts. To achieve this goal, we develop feature selection approaches to select climate variables of certain time ranges, and an efficient predictive model called the generalized parallel partial Gaussian process (G-PPGP) to capture the association between the selected features  and peak NDVI. This model achieves higher accuracy than  pixel-wise linear regression \citep{williams2023anthropogenic} and two other deep learning models. 

\subsubsection{Feature Selection }\label{subsec:feature_selection}
Past research \citep{williams2023anthropogenic} found winter-to-summer precipitation and summer VPD as key drivers of NDVI variability in the Four Corners region. The August NDVI is primarily influenced by boreal spring
 soil moisture levels, which are affected by the precipitation and VPD values in the previous months leading up to August. High August VPD values can also cause plants to shut their stomates and reduce photosynthesis, which inhibits vegetation growth. 

 To test which temporal windows of these climate variables are most informative for predicting August NDVI, we use two feature selection approaches. The first method follows \cite{williams2023anthropogenic} and uses linear models, where the August NDVI is the response variable and various monthly ranges of VPD and precipitation are used as covariates. We found that the average July-August VPD and January-August precipitation are the most informative with the largest $R^2$ (defined in Equation \eqref{grand_r2} of Appendix \ref{variable_selection}), consistent with what is found in \cite{williams2023anthropogenic}. The second approach is non-linear and consists of computing the maximized marginal likelihoods of various combinations of monthly ranges. This method also resulted in July-August VPD and January-August precipitation as the most informative covariates. Detailed results for both  approaches are provided in Appendix \ref{variable_selection}. In the remaining sections, we denote August NDVI values as $y_{i}(t_j)$ for the year $t_j$ at spatial grid $i$, and the climate attributes as $x_{\text{precip},i}(t_j)$, $x_{\text{vpd},i}(t_j)$ representing the January-August average precipitation and July-August average VPD, respectively.

\subsubsection{Attribution Models (Phase I)}\label{subsec:attribution}

We first develop an attribution model that uses observed covariates from the prediction year to estimate how NDVI responds to precipitation and VPD. This step assures that both the selected variables and the chosen model structure can accurately represent the relationship between NDVI and covariates before applying the method to long-term forecasting. As the relationship may not be linear, it motivates us to consider Gaussian processes (GPs)  \citep{rasmussen2006gaussian}, which are widely used for predicting nonlinear relationships with quantified uncertainty. A straightforward approach is to build a pixel-wise Gaussian process at each spatial location. However, training GPs with distinct correlation parameters separately at each location is computationally expensive and unstable. To avoid this limitation, the parallel partial Gaussian process (PPGP) was developed in \cite{gu2016parallel}, which has distinct mean and variance parameters, yet the correlation matrix is shared across all grids. 
PPGP is much faster and more stable than separable GPs because the numerous mean and variance parameters can be computed in closed form, and only one set of correlation parameters in the correlation matrix needs to be numerically optimized.   

We extend this framework to develop the G-PPGP, in which the output at each spatial grid still shares the same correlation parameters, but the correlation matrix differs across grids due to the different inputs, such as precipitation and VPD. Allowing for unique inputs per grid captures  the spatial variability of these covariates because precipitation and VPD conditions differ significantly across the region.

For each grid $i= 1, \ldots, k$, the NDVI at that location can be modeled as $y_i(\mathbf{x}) = f_i(\mathbf{x}) + \epsilon_i$, where $y_i$ is the August NDVI at location $i$, $f_i(\mathbf{x})$ is a GP  at climate variables $\mathbf{x} \in \mathbb R^p$ with a constant mean parameter $\mu_i \in \mathbb R$, a covariance function $\sigma_i^2 K(\cdot,\cdot)$,  and $\epsilon_i\sim N(0,\eta \sigma^2_i)$ being an independent Gaussian noise often attributed to the interval variability of the observables \citep{deser2012uncertainty}. In our application, $\mathbf x$ is  the January-August precipitation average and July-August VPD average of a certain year at location $i$, and thus $p=2$. 

Suppose we have $n$ years of past data, leading to any $p \times n$ input matrix $\mathbf{X}_i =[\mathbf{x}_i^{T}(t_1),\ldots, \mathbf{x}_i^{T}(t_n)]$, integrating out the latent Gaussian process $f_i(\cdot)$, the marginal distribution of $\mathbf{y}_i = [y_i(\mathbf{x}_i(t_1)), ... , y_i(\mathbf{x}_i(t_n))]$ follows a multivariate normal distribution 
\begin{equation}\label{general_ppgp}
\left(\mathbf{y}_i \mid \mathbf{X}_i, {\mu}_i, \sigma_i^2, \bm \gamma, \eta \right) \sim \mathcal{M N}\left({\mu}_i \mathbf 1_n, \sigma_i^2 \Tilde{\mathbf{K}}_i\right),
\end{equation}
where $\mathbf 1_n$ is an $n$-vector of ones, and $\Tilde{\mathbf{K}}_i = \mathbf{K}_i + \eta \mathbf{I}_n$ is the covariance  with $\mathbf{K}_i$ being a correlation matrix, $\eta$ being the nugget parameter, and $\bm \gamma= (\gamma_{1},\ldots,\gamma_{p})$ being the correlation or range parameters. The $(t,t')$-th entry of $\mathbf{K}_i$ follows $K(\mathbf{x}_i(t),\mathbf{x}_i(t'))$ where $K(\cdot,\cdot)$ is a kernel function. We use a product of the correlation function at each coordinate \citep{sacks1989design,Gu2018robustness}, as the inputs have distinct physical meaning, 
\[K(\mathbf{x}_i(t),\mathbf{x}_i(t')) = \prod_{l=1}^{p}K_{l}(x_{l,i}(t), x_{l,i}(t^{\prime})).\]
For each $K_l$, we use the Mat{\'e}rn covariance with roughness parameter being 2.5, which is the default kernel used in GP packages \cite{roustant2012dicekriging,gu2018robustgasp}:  
\begin{equation}\label{matern}
K_{l}(x_{l,i}(t), x_{l,i}(t^{\prime}))=\left(1+\frac{\sqrt{5}d}{\gamma_{l}}+\frac{5 d^2}{3 \gamma_{l}^2}\right) \exp \left(-\frac{\sqrt{5} d}{\gamma_{l}}\right),
\end{equation}
where $d = |x_{l,i}(t)- x_{l,i}(t^{\prime})|$ for $l=1,\ldots,p$. 
As the parameters $\mu_i$ and $\sigma^2_i$ can be integrated out explicitly for computing the predictive distribution of a future year, for $i=1,...,k$, only $p+1$ parameters $\{\bm \gamma, \eta\}$ need to be numerically estimated. We discuss the estimation of $\eta$ and $\bm \gamma$  in Section \ref{sec:hyperparam} of the Appendix.

Given the training data and the  January-August precipitation and July-August VPD of location $i$ of a future year $t_{*}$, denoted as $\mathbf x_{i}(t_{*})$, 
following \cite{gu2016parallel}, we  integrate out the mean and variance parameters. Thus, the predictive distribution of NDVI for year $t_{*}$ at location $i$ follows a non-central Student's $t$-distribution with $n-1$ degrees of freedom:
\begin{align}\label{eq-gp-pred}
    (y_{i}(\mathbf x_i(t_{*}))\mid\mathbf X_{i}, \mathbf y_{i}, \mathbf x_i(t_{*}), \bm \gamma, \eta)\sim \mathcal{T}(\hat{y}_{i}(\mathbf x_i(t_{*})), \hat{\sigma}_{i}^{2}K_{i}^{*}, n-1),
\end{align}
where the predictive mean and scale parameters follow
\begin{align}
    \hat{y}_{i}(\mathbf x_i(t_{*})) & = \hat{\mu}_{i} + \mathbf k^{T}(\mathbf x_i(t_{*}))\Tilde{\mathbf K}_{i}^{-1}(\mathbf y_{i}-\hat{\mu}_{i}\mathbf{1}_{n}),\label{eq:pred_mean}\\
    \hat{\sigma}_{i}^{2} & = \frac{1}{n-1}(\mathbf y_{i}-\hat{\mu}_{i}\mathbf{1}_{n})^{T}\Tilde{\mathbf K}_{i}^{-1}(\mathbf y_{i}-\hat{\mu}_{i}\mathbf{1}_{n}),\label{eq:pred_var}\\
    K^{*}_{i} & = 1+\eta-\mathbf{k}^{T}(\mathbf x_i(t_{*}))\Tilde{\mathbf K}^{-1}_{i}\mathbf{k}(\mathbf x_i(t_{*}))+\frac{\left(1-\mathbf{1}^{T}_{n}\Tilde{\mathbf K}_{i}^{-1}\mathbf{k}(\mathbf x_i(t_{*}))\right)^{2}}{\mathbf{1}^{T}_{n}\Tilde{\mathbf K}_{i}^{-1}\mathbf{1}_{n}}.\label{eq:pred_cov}
\end{align}
Here, $\hat{\mu}_{i} = \left(\mathbf{1}_{n}^{T}\Tilde{\mathbf K}_{i}^{-1}\mathbf{1}_{n}\right)^{-1}\mathbf{1}_{n}^{T}\Tilde{\mathbf K}_{i}^{-1}\mathbf{y}_{i}$ is the generalized least square estimator of the mean, 
and 
$\mathbf k(\mathbf x_i(t_{*}))=(K(\mathbf x_i(t_1),\mathbf x_i(t_{*})),\ldots,K(\mathbf x_i(t_n),\mathbf x_i(t_{*})))^{T}$
is an $n$-vector representing the covariance between the training  inputs and the test input at time $t_{*}$. The predictive mean $  \hat{y}_{i}(\mathbf x_i(t_{*}))$ is often used for prediction, and the uncertainty can be quantified from the closed-form expression of the intervals in the predictive distribution. 

In addition to the G-PPGP attribution model, we test a pixel-wise linear regression model and two popular deep learning methods: a feedforward Deep Neural Network (DNN) \citep{svozil1997introduction} and a Fourier Neural Operator (FNO) \citep{li2020fourier}. The pixel-wise linear regression model is equivalent to the model studied in  \cite{williams2023anthropogenic}, where the inputs are also July-August VPD and January-August precipitation up to and including year $t$ for each grid. The mathematical details of this model are in Appendix \ref{sec:hyperparam}. For the DNN model, similar to the pixel-wise linear regression model in Equation \eqref{eq-lm-attr}, we train a separate DNN for each location, where each network takes a vector of precipitation and VPD for the year $t_j$ at location $i$. For 
FNO, spatial maps of precipitation and VPD at time $t_j$ are provided as inputs to predict NDVI at year $t_j$. We use the activation function rectified linear unit (ReLU) for both methods, which is defined for each element as $\mbox{ReLU}(x)=\mbox{max}(0,x)$. Background information and model details about DNN and FNO are in Appendix \ref{sec:hyperparam}, with hyperparameters and training details 
in Tables \ref{dnn_param} and \ref{fno_param-a} of the Appendix. 

\subsection{Forecasting Climate Attributes and Capturing Dynamic Correlation (Phase II)}\label{subsec:one-year-ahead_cov}

With the attribution model established, we proceed to the next stage of the framework, where the climate covariates are forecasted so that the model no longer relies on information from the prediction year. In this section, we analyze the temporal autocorrelation of the covariates and develop appropriate models for forecasting both precipitation and VPD.

\begin{figure}[H]
    \centering
\includegraphics[width=1\textwidth]{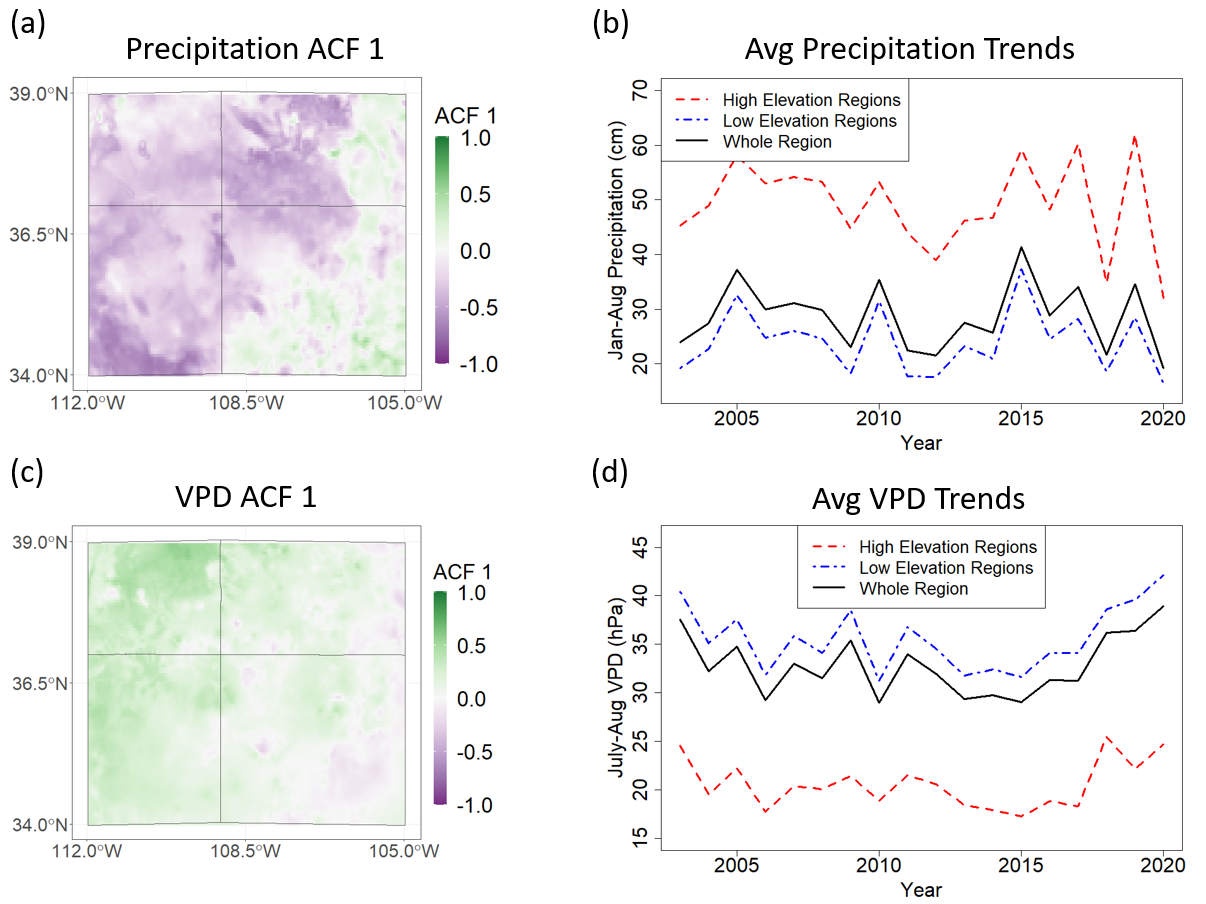}
    \caption{Panels (a) and (c): The lag 1 autocorrelation values for each location between the years 2003-2020 for precipitation and VPD. Panels (b) and (d): Annual precipitation and VPD averages over high elevation regions ($\geq$ 2500 m), low elevation regions ($<$ 2500 m), and the whole region (low and high elevation regions).}
    \label{fig:precip_volatility}    
\end{figure}

In the Four Corners region of the United States during the years 2003 to 2020, we found that January-August precipitation measurements tend to be negatively correlated with their previous values for the majority of the spatial grids, while we found the July-August VPD measurements at most grids are positively correlated with the values from the previous year; the positive VPD correlation is tied to the influence of increasing air temperatures \citep{williams2023anthropogenic}. Figure  \ref{fig:precip_volatility}(a) and \ref{fig:precip_volatility}(c) depict the autocorrelation function at lag 1 (ACF 1) of precipitation and VPD, respectively. The mean of the precipitation ACF 1 values is -0.207, while the mean of the VPD ACF 1 values is 0.155.  As the average ACF 1 for a large number of white noise processes of 18 time points is approximately $-0.056$ \citep{kendall1954note,nickell1981biases}, these results indicate, on average, precipitation has a negative autocorrelation and VPD has a positive autocorrelation. Furthermore, Figure \ref{fig:precip_volatility}(b) shows that the average January-August precipitation amounts from year to year often jump between high and low values, and these jumps may be associated with climate extremes \citep{donat2016more,chang2016changes,li2022evaluation}. On the other hand, Figure \ref{fig:precip_volatility}(d) shows that VPD is a lot less volatile, and strong secular increases in VPD may be predicted based on the values from the previous years. Therefore, it is crucial that correlation parameters for VPD and precipitation are uniquely determined.

To model precipitation and VPD, we build an  autoregressive model of order 1 
\citep{West1997}, which enables both positive and negative temporal correlation patterns to be captured.   Following the PPGP model \citep{gu2016parallel}, we assume the correlation of the covariates is shared across spatial locations, which substantially reduces the computation. 
The correlation of precipitation follows $\mbox{Cor}(x_{\text{precip}, i}(t),  x_{\text{precip}, i}(t'))=\rho_{\text{preci}}^{|t'-t|}$, where $\rho_{\text{preci}} \in (-1,1)$.
 Here, $\rho_{\text{preci}}$ captures the correlation of precipitation at a location between two consecutive years.  As $\rho_{\text{preci}}$ is estimated by data, it enables the model to capture the negative correlation patterns from the data. The correlation of VPD is assumed to be the Mat{\'e}rn correlation in Equation (\ref{matern}), which implicitly imposes non-negative correlation for VPD over time.  We use cross-validation to estimate the hyperparameters in this model, as detailed in Section \ref{sec:forecast_hyperparam} of the Appendix.

 In addition to PPGP, we also implement linear models, FNO, historical averages, and a previous-year baseline (using the prior year’s observation as the prediction) to forecast covariates. These results are also discussed in detail in Appendix \ref{sec:forecast_hyperparam}.

\subsection{Long-Term NDVI Forecasting (Phase II)}\label{ndvi_forecast}

Once precipitation and VPD are predicted one year in advance, these predictions can be used as additional input information to forecast NDVI one year into the future. Using model~\eqref{general_ppgp} with $n$ training years, we forecast the August NDVI of location $i$ at year $t_{n+1}$ with model~\eqref{eq-gp-pred}, but replace the input $\mathbf x_{i}(t_{n+1})$ with the predicted covariates $\hat{\mathbf x}_{i}(t_{n+1}) = (\hat{x}_{\text{precip},i}(t_{n+1}), \hat{x}_{\text{vpd},i}(t_{n+1}))^{T}$.
Additionally, for comparison, we constructed a linear model where the NDVI output follows Equation (\ref{eq-lm-attr}), but the inputs are linearly forecasted VPD and precipitation.

In addition to GPs and linear models for one-year-ahead forecasting, we also use FNO to predict August NDVI one year ahead. We use a similar architecture to the FNO model in Section~\ref{subsec:attribution}, except that the input matrix is the spatial map of NDVI in the previous year.
The information pertaining to  the number of layers, neurons, and structure of the forecast specific FNO is in Table \ref{fno_param-f} of the Appendix. Lastly, we also compare it with a recurrent neural network (RNN) model \citep{medsker1999recurrent} as a NDVI long-term forecasting neural network benchmark, as it is designed to model sequential data. We train the RNN model for each grid separately, similar to the DNN attribution model in Section \ref{subsec:attribution}. For detailed specifications regarding the structure, input and output dimensions, the number of epochs, and other parameters specific to our RNN application, refer to Table \ref{rnn_param} in the Appendix.

\subsection{Performance Metrics and Uncertainty Quantification}

To evaluate the predictive performance of each model, we calculate the root mean square error (RMSE), 95\%  predictive credible interval (CI), proportion of observations covered by the 95\% predictive CI, and length of the CI:
\begin{equation*}
\begin{aligned} 
\text{RMSE} & = \left(\frac{1}{kn^{*}} \sum_{i=1}^k\sum_{j=n+1}^{n+n^{*}}\left(y_i\left(t_j\right)-\hat{y}_i\left(t_j\right)\right)^2\right)^{\frac{1}{2}},
\\ P_{95 \% }& =\frac{1}{k n^{*}} \sum_{i=1}^k \sum_{j=n+1}^{n+n^{*}} 1_{y_i\left(t_j\right) \in \mbox{CI}_{i, 95 \%}(t_j)}, 
\\ L_{95 \%} & =\frac{1}{k n^{*}} \sum_{i=1}^k \sum_{j=n+1}^{n+n^{*}} \operatorname{length}\left\{\mbox{CI}_{i, 95 \%}(t_j)\right\},\end{aligned}
\end{equation*}
where $\hat{y}_i(t_j)$ is the predicted output (NDVI, precipitation, or VPD, depending on which is being modeled) at location $i$ and time $t_j$, with $n$ and $n^{*}$ being the number of years in the training data and held-out test data, respectively. Here, $P_{95\%}$ is the proportion of true observed values that are covered by the 95\% CI. The $L_{95\%}$ is the average length of the CI over all locations and prediction years. A method with a small RMSE, $P_{95\%}$ close to the nominal $95\%$ level, and a small $L_{95\%}$ is ideal for predictions and uncertainty quantification. 

\section{Results}\label{results}

\subsection{Attribution Models}
Table \ref{tab:attribution_results} compares different attribution models for predicting peak (August) NDVI introduced in Section \ref{subsec:attribution}, where average July-August VPD and average January-August precipitation, from the prediction year, are used as the input variables. 
We also include the location mean as a baseline model, which is the pixel-wise linear model with only the intercept as a covariate. The first 10 years from 2003 to 2012 are used as the training data, and data from the last 8 years from 2013 to 2020 are held out as test data. 
The G-PPGP model outperforms other alternatives in terms of predictive RMSE. Furthermore, the assessed uncertainty by G-PPGP is precise, as around $95\%$ of the held-out test data are covered in a short $95\%$ posterior credible interval. 

\begin{table}[H]
\caption{Comparison between attribution models for the out-of-sample prediction of NDVI by using July-August average VPD and January-August precipitation, from the prediction year, as the inputs. The RMSE for all methods are reported. The proportion of the held out data covered by the 95\% CI  and the average length of the 95\% CI are reported for the G-PPGP and LM methods. The best values are bold. 
}	
		\label{tab:attribution_results}
		\centering
		\begin{tabular}{lccccc}
			\hline
	     Precip and VPD Known & G-PPGP & LM & DNN & FNO & 
         Location Mean\\
			\hline  
			  RMSE & {\bf 0.0291}  &  0.0305 & 0.1728 & 0.0421 & 
              0.0363\\
              $P_{95\%}$ & {\bf 0.949} &  0.938 & --- & --- & 
              ---\\
              $L_{95\%}$ & {\bf 0.110} &  0.113  & --- & --- & 
               ---\\
              
			\hline
		\end{tabular}
\end{table}

\begin{figure}[H]
    \centering
\includegraphics[width=.9\textwidth]{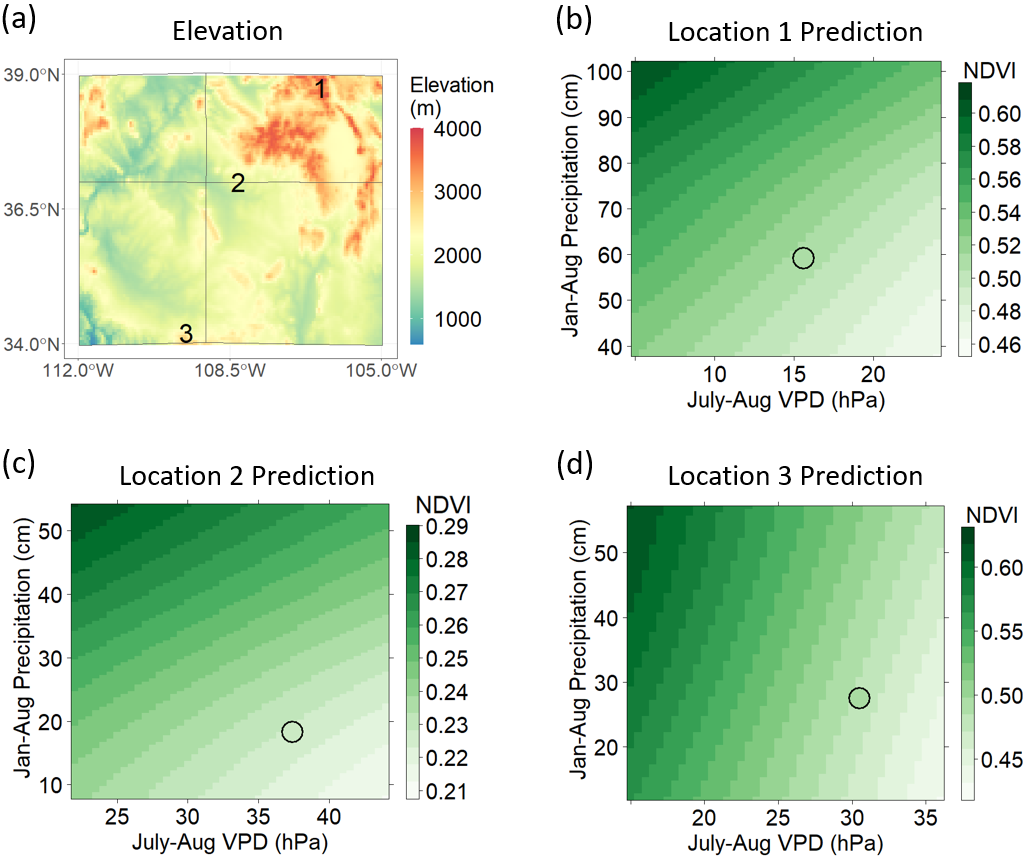}
    \caption{ 
    (a) Elevation map with three locations marked as "1"-"3".  Panels (b)-(d) show the heatmaps  of attribution predictions of August NDVI by the G-PPGP model given various January-August precipitation and July-August VPD values at locations 1-3, respectively. The circle locations represent the recorded precipitation and VPD in 2020, and the inside color of the circles represents the true August NDVI value observed that year. 
    }
    \label{fig:elevation_grid_attribution}
\end{figure}

Panels (b), (c), and (d) of Figure \ref{fig:elevation_grid_attribution}  show the attribution predictive mean of G-PPGP for the year 2020 at three separate locations that experience different precipitation and VPD conditions. The difference between each heatmap is due to other confounding factors, such as elevation and terrain at each location, which are not included as covariates in the regression models, as they are invariant over the years. 
 Additionally, for each prediction heatmap, a circle is plotted to indicate the real average January-August precipitation and July-August VPD for the year 2020. The inner color of the circle indicates the true NDVI observed during the year 2020 at that location, and they are similar to the predictive values from G-PPGP. 
 We plot maps of the true August NDVI and attribution-predicted August NDVI of years 2019 and 2020 in Figure \ref{fig:2019_2020_attribution} of Section \ref{sec:add_att_models} of the Appendix.

\subsection{One-Year-Ahead Forecasts of NDVI}\label{subsec:one-year-ahead}

Prior to predicting NDVI, we inspect the predictive accuracy of precipitation and VPD in Section \ref{cov_forecast_supp} 
of the Appendix. Table \ref{tab:precip_forecast} 
contains performance metrics to compare different methods of forecasting precipitation.
 The PPGP method outperforms all methods with the smallest RMSE because it captures correlation patterns of climate attributes, such as the negative correlation of precipitation discussed in Section \ref{subsec:one-year-ahead_cov}. 
The forecast performance of July-August VPD is provided in Figure \ref{fig:vpd_results} 
and Table \ref{tab:vpd_forecast} 
in Appendix \ref{cov_forecast_supp}. Out of all the methods, PPGP is the most robust method with consistently good performance. The reliable forecast of climate variables improves the accuracy of the long-term forecast of peak NDVI.

As for NDVI, we first consider the one-year-ahead forecast of gross August NDVI, defined to be the sum of NDVI at all spatial grids in August. 
Table \ref{tab:gross_ndvi_residual_table} gives the RMSE and residuals of gross August NDVI forecasts from different methods described in Section \ref{ndvi_forecast}. All methods use data at least one year before the month of interest for the one-year-ahead forecast, and the inputs of each model are provided in Table \ref{tab:best_model_results}. Provided the RMSE and residuals of the gross August NDVI forecasts, the G-PPGP approach substantially outperforms all other models. 

\begin{table}[H]
		\caption{RMSEs of the one-year-ahead forecast of the August gross NDVI  and the  residual values (true minus forecasted gross August NDVI) for each year. The values closest to 0 are bold, as they indicate the best performance.}	
		\label{tab:gross_ndvi_residual_table}
		\centering

        \begin{tabular}{lcccccccccc}
            \hline 
           
	Method &RMSE & 2013 & 2014 & 2015 &  2016 & 2017 & 2018 & 2019 & 2020\\
			\hline
        Location Mean  &250 & 114 & 69.7 & 406 & 138 & 214 & -374 & -23.2 &  -337\\
        Previous Yr & 337 & 298 & -37.9 & 370 & -257 & \textbf{91.0 }& -613 & 349 & -333\\
        FNO & 263 & 170 & 39.6 & 409 & 59.1 & 202 & -439 & 124 & -324\\
        RNN & 288 & 305& 109  & 397 &234  &250  &-351  & \textbf{14.4} & -402 \\
        AR(1)& 268 & 111 & 96.0 & 463 & 210 & 282 & -348 & 40.7 & -306\\
        LM  & 291 & 148& 56.3 &406 & \textbf{1.24 }& 94.1 & -556  &  -91.0 & -404\\
	    G-PPGP& {\bf 198} & \textbf{101} &\textbf{22.6} &\textbf{365} & 84.0 & 110 & \textbf{-345} & -51.8 & \textbf{-174}\\
	      
			\hline
		\end{tabular}
	\end{table}

\begin{figure}[H]
    \centering
    \includegraphics[width=1\textwidth]{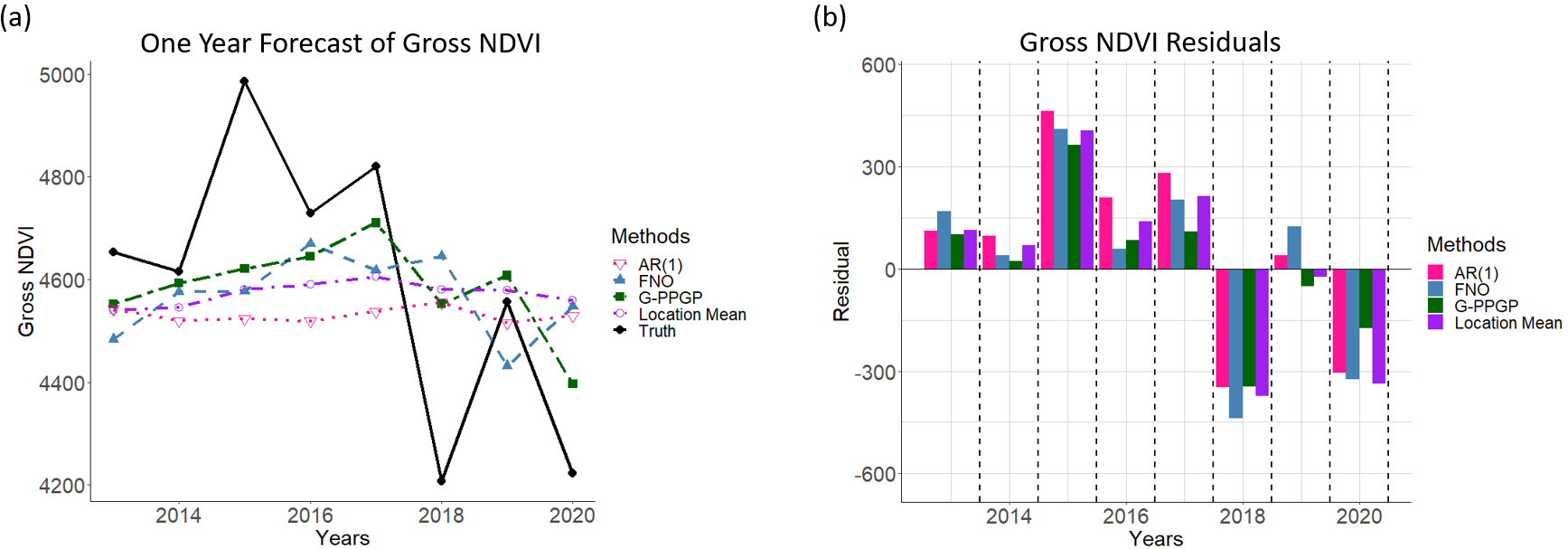}
    \caption{(a) The true gross NDVI, calculated as the sum of all August NDVI values over all grids for each prediction year, and the one-year-ahead forecasted gross NDVI using the best four methods. (b) The residuals between true gross NDVI and forecasted gross NDVI for the four methods.}
    \label{fig:overall_ndvi_forecast}
\end{figure}

To explore the pattern of the one-year-ahead forecast of each method, 
we plot the true gross NDVI and predicted gross NDVI using the four most accurate models in Figure \ref{fig:overall_ndvi_forecast}(a). The green dashed line  is the predicted gross NDVI using G-PPGP, which best captures the trend 
of the true gross NDVI. 
A notable decreasing trend of NDVI after 2016, which is associated with the increase of VPD during this period shown in Figure \ref{fig:data_image}(d), is correctly predicted by G-PPGP. The learned relationship between NDVI and VPD one year ahead enables the reliable prediction of NDVI. Additionally, the G-PPGP method picks up fluctuations in NDVI better than other models, such as the first-order autoregressive model (AR(1)), the location mean, and FNO. Figure \ref{fig:overall_ndvi_forecast}(b) plots the residuals, the difference between the true gross NDVI and predicted gross NDVI for these four methods, and Figure \ref{fig:full_gross_ndvi}  in the Appendix 
shows the results of all the methods. 
Overall, the G-PPGP has the smallest RMSE, and it has the smallest absolute residuals for five out of the eight prediction years among all models.    

Furthermore, the one-year-ahead NDVI forecast results for each spatial grid are provided in Table \ref{tab:best_model_results}. The best model with the smallest RMSE is the G-PPGP method, which takes the predicted January-August precipitation and  July-August VPD as input. The improvement in performance of the G-PPGP method in Table \ref{tab:best_model_results} is not as large as the forecast of gross NDVI in Table \ref{tab:gross_ndvi_residual_table} due to the large variability in the gridded data.
All methods that produce uncertainty quantification in Table \ref{tab:best_model_results} are slightly overconfident as the percentage of held-out test data is less than 95\%. This is due to the plug-in estimates of the predicted inputs; the uncertainty quantification can be improved by propagating the uncertainty through the posterior predictive samples of climate attributes instead of using the plug-in estimates. 
Furthermore, the proposed G-PPGP method significantly outperforms the deep learning methods, FNO and RNN,  as deep learning methods often require a large amount of training data.   

In addition to predicting one year into the future, Section \ref{sec:less_than_one_yr} in the Appendix 
provides an extension of the G-PPGP model that uses half-year-ahead to one-month-ahead data to forecast August NDVI. The forecast becomes more accurate when using more information closer to the month to be predicted.  

\begin{table}[H]
		\caption{Model performance of the one-year-ahead forecasts for each spatial grid and all test years. The Location Mean method is the per-grid August NDVI average over the past known years. The Previous Yr method uses last year's NDVI at each location as the forecast. FNO is the Fourier Neural Operator method. RNN is the Recurrent Neural Network. AR(1) is an autoregressive model of order 1.  LM is a linear model where the one-year-ahead forecast of  precipitation and VPD, via linear models, is plugged into the model as inputs. G-PPGP is the generalized PPGP method, where the one-year-ahead forecast of precipitation and VPD, by PPGPs, is used as input. The inputs for each model are specified in the input column. "Historical NDVI" is all NDVI data before the year of prediction (using $t_j-1$ years of data to predict year $t_j$). The best values
are bold.
  }	
		\label{tab:best_model_results}
		\centering
		\begin{tabular}{lcccc}
			\hline
	    Method & Input & RMSE & $P_{95\%}$ & $L_{95\%}$ \\
			\hline  
   Location Mean & Historical NDVI & 0.0363 & --- & --- \\
Previous Yr & Historical NDVI &0.0474 & --- & --- \\

FNO & Historical NDVI & 0.0399 & --- & ---
\\
RNN & Historical NDVI, VPD, and Precip. & 0.0435 & --- & ---

\\
AR(1) & Historical NDVI & 0.0368 &  {\bf 0.899} & 0.117
\\
 LM  & LM Predicted VPD and Precip. &0.0393 & 0.835& 0.105
\\
G-PPGP & PPGP Predicted VPD and Precip. & {\bf 0.0352} & 0.883 &{\bf 0.103}
  
            \\
			\hline
		\end{tabular}
\end{table}

\begin{figure}[H]
    \centering
    \includegraphics[width=1\textwidth]{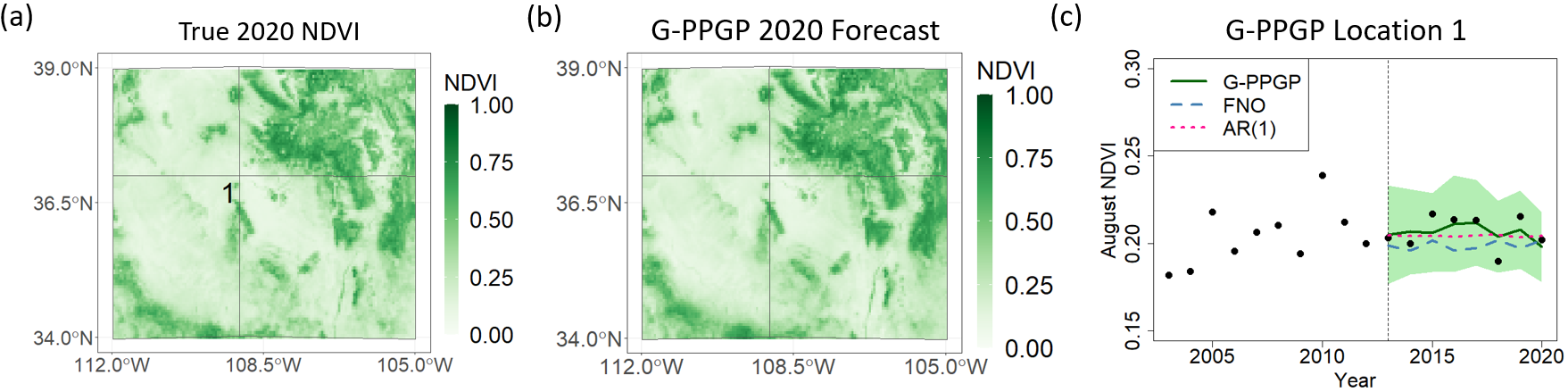}
    \caption{(a) The true August NDVI in the year 2020. (b) The forecast of  August NDVI using the G-PPGP method with the predicted precipitation and VPD, by PPGP, as inputs for the year 2020. (c) G-PPGP, FNO, and AR(1) yearly forecasts for the test years. The light green area is the 95\% CI for location "1" using G-PPGP.}
    \label{fig:ndvi_results}
\end{figure}

 The true August NDVI and G-PPGP prediction for the year 2020 are provided in Figure \ref{fig:ndvi_results}(a) and Figure \ref{fig:ndvi_results}(b), respectively. Figure \ref{fig:ndvi_results}(c) plots the forecasts using G-PPGP, FNO, and AR(1) for a specific location "1". In this location, FNO  consistently underestimates high-value August NDVI years, and AR(1) seems to  model close to a flat line with no fluctuations. Compared to these two models, G-PPGP is better at capturing yearly volatility.  
It is also important to note that the G-PPGP model with forecasted inputs performs significantly better than using only historical $t_j-1$ years of data as inputs (Table \ref{tab:prev_vdp_prev_precip_results} of Appendix \ref{additional_appendix_models}).

\section{Discussion}\label{discussion}

In this work, we built a two-phase model for the long-term forecast of NDVI. In phase one, we develop a model that best captures the relationship between climate variables, including precipitation and VPD, and peak NDVI, using climate inputs from the same year as the NDVI being predicted. In phase two, we build a dynamic linear model that captures both positive and negative correlations of the climate attributes, and provide a one-year-ahead forecast for NDVI based on the predicted climate variables.
The forecasts from the two-phase approach developed in this work are not only more accurate but they are also more interpretable to ranchers and farmers, compared to conventional black-box machine learning models. 
The improvement in performance of the one-year-ahead NDVI forecast using our two-phase method is due to three reasons. 
First, the grid-based peak NDVI is highly volatile between years, and directly building predictive models for NDVI leads to inaccurate predictions between two consecutive years. To address this issue, we found that the correlation between climate attributes and NDVI is strong; VPD was found to reduce global NDVI \citep{yuan2019increased}, while on the other hand, precipitation is often positively correlated with NDVI. An NDVI forecast model conditional on reliable forecasts of VPD and precipitation for future years can be more accurate than a direct one-year-ahead forecast of NDVI.
Second, we found that VPD is positively associated with values from previous years, whereas precipitation shows a negative association with values from previous years in the Four Corners region. These correlation patterns are captured by our model, which enhances the accuracy of NDVI forecasting.
Third, the generalized parallel partial Gaussian process (G-PPGP) developed in this work requires far fewer yearly data points than other machine learning approaches, such as deep neural networks and operators, as it contains significantly fewer parameters that require numerical optimization.
Our method is applicable to other climate datasets with limited yearly data that exhibit complex correlation patterns.

The majority of machine learning approaches for forecasting NDVI focus on short-term predictions, such as one-month-ahead forecasts. This article fills the gap by providing a one-year-ahead actionable forecast of NDVI with quantified uncertainty through a two-phase model that is interpretable for farmers and policymakers. Compared to Grass-Cast \citep{hartman2020seasonal}, we provide forecasts of peak NDVI on spatial grids with higher resolution and do not require users to input climate conditions for future years.
Furthermore, grid-based NDVI forecasting requires selecting informative climate variables over a given time range. The climate variables selected by our approach are consistent with the findings of \cite{williams2023anthropogenic}, which developed linear regression models between the selected climate variables and the NDVI outcome. To capture the nonlinear association between climate variables and peak NDVI, we built computationally scalable Gaussian process models that improve forecast accuracy.


The correlation patterns of climate variables between consecutive years are worth further exploration. It is well known that precipitation is positively correlated with NDVI and that rainfall lag is informative for predicting vegetation amount \citep{ji2003assessing, piao2003interannual, potter1999interannual}. Furthermore, NDVI can also be highly sensitive to precipitation anomalies \citep{robinson2021climate, gessner2013relationship}. Extreme precipitation events such as floods and droughts are becoming more frequent and severe worldwide \citep{donat2016more, allen2002constraints}, which likely explains the negative autocorrelation we observed for precipitation in the Four Corners region, as discussed in Section \ref{subsec:one-year-ahead_cov}. Consequently, in an increasingly volatile climate, the proposed G-PPGP model performs well because it can capture both positive and negative correlations, and its correlation parameter is updated as more data becomes available.

\section{Conclusions}\label{conclusion}

We developed a machine learning framework that first forecasts informative climate attributes, including precipitation and VPD, and then predicts vegetation conditions based on the forecasted climate attributes one year into the future. This framework was applied to the Four Corners region of the United States. Our models effectively capture distinct correlation patterns of precipitation and VPD to predict these climate attributes.
We constructed a generalized parallel partial Gaussian process and demonstrated its substantial improvement in one-year-ahead forecasts of gross peak NDVI compared with other alternatives.

Long-term forecasts of climate and vegetation conditions remain a crucial task. Such forecasts can inform important decisions, including those related to buying or selling livestock. Furthermore, the approach employed here is straightforward for end users to understand and update. For example, if precipitation from January to May has been abundant and contradicts the forecast made the previous August, users can adjust their outlooks accordingly.

Several directions are worth exploring. First, when forecasting NDVI, one can propagate the uncertainty from forecasting climate attributes through posterior predictive samples to improve the uncertainty quantification of vegetation forecasts. Second, due to the large input space and increasing volatility of climate variables, directly incorporating variable selection as part of the modeling process could improve accuracy and quantification of uncertainty. Third, global patterns, such as gradients of ocean surface temperatures and ensembles of climate simulations, may provide additional information for achieving more accurate long-term forecasts. Lastly, it is of great interest to relate forecasted vegetation conditions to crop growth and biomass production, evaluated by above-ground net primary production, and to expand the study to other regions to obtain reliable forecasts that can enhance food security. 



\vspace{6pt}

\funding{We acknowledge the support from the National Science Foundation under Award No. ITE-2236021. 
}

\dataavailability{The data and code used in this work are publicly available (\url{https://github.com/UncertaintyQuantification/forecast_ndvi}).
}

\conflictsofinterest{The authors declare no conflicts of interest.
} 


\newpage
\abbreviations{Abbreviations}{
The following abbreviations are used in this manuscript:
\\

\noindent 
\begin{tabularx}{\linewidth}{@{}p{2cm}X@{}}
NDVI & Normalized Difference Vegetation Index\\
U.S. & United States\\
ANPP & Aboveground net primary production\\
VPD & Vapor pressure deficit\\
ARIMA & Autoregressive integrated moving average\\
LM & Linear Model\\
G-PPGP & Generalized parallel partial Gaussian process\\
MODIS &  Moderate Resolution
Imaging Spectroradiometer\\
USGS & United States Geological Survey\\
PRISM & Parameter-elevation Regressions on Independent Slopes Model\\
LSTM & Long Short-Term Memory\\

GP & Gaussian process\\
PPGP & Parallel partial Gaussian process\\
DNN & Deep Neural Network\\
FNO & Fourier Neural Operator\\
ReLU & Rectified Linear Unit\\
ACF & Autocorrelation Function \\
RNN & Recurrent Neural Network \\
RMSE & Root Mean Square Error\\
CI & Credible Interval\\
AR & Autoregressive \\
3D & 3-dimensional\\
MLE & Maximum Likelihood Estimator\\

\end{tabularx} }

\appendixtitles{no} 
\appendixstart
\appendix
\section[\appendixname~\thesection]{}
\subsection[\appendixname~\thesubsection]{Temporal Window Feature Selection for Climate Attributes}\label{variable_selection}
\textbf{Pixel-Wise Linear Regression.} 
As each pixel of the data represents a spatial grid, the pixel-wise linear regression model at each grid $i$ can be written as
\begin{equation}\label{eq-lm-attr}
    y_i(t_j)= \mathbf z_i(t_j)\bm \beta_{i} +\epsilon_{i}(t_j),
\end{equation}
where $y_i(t_j)$ is the NDVI at location $i$ and time $t_j$, and $\epsilon_{i}(t_j)$ denotes Gaussian white noise with variance $\sigma_{i}^{2}$. Here, $i=1,\ldots,k$ and $j=1,\ldots,n$ where $k$ is the number of grids and $n$ is the number of years. The covariates $\mathbf z_i(t_j)=(1,x_{\text{precip},i}(t_j),x_{\text{vpd},i}(t_j))$ is a 3-dimensional (3D) row vector of the intercept, precipitation, and VPD at grid $i$ and year $t$, and $\bm \beta_{i}=( \beta_{0,i},\beta_{\text{precip},i},\beta_{\text{vpd},i})^T$ is the corresponding 3D vector of parameters.  Next, we denote $\mathbf y_i=(y_i(t_1),\ldots,y_i(t_n))$ and $\mathbf z_i=(\mathbf z^T_i(t_1),\ldots, \mathbf z^T_i(t_n))^T$ which are an $n$ vector of outcomes and an $n\times 3$ matrix of covariates, respectively. Separately for each grid $i$, $i=1,\ldots,k$,  the maximum likelihood estimator (MLE) of the parameters follows $\bm {\hat \beta_{i}}=(\mathbf z^T_i \mathbf z_i)^{-1}\mathbf z^T_i \mathbf y_i$.  For any new covariate $\mathbf z_i(t_{*})$, the prediction for year $t_*$ can be obtained by plugging in the estimated parameter $\mathbf{\hat y}^*_i= \mathbf z_i(t_{*}) \hat{\bm \beta_i}$. 

\begin{figure}[H]
    \centering
    \includegraphics[width=.9\textwidth]{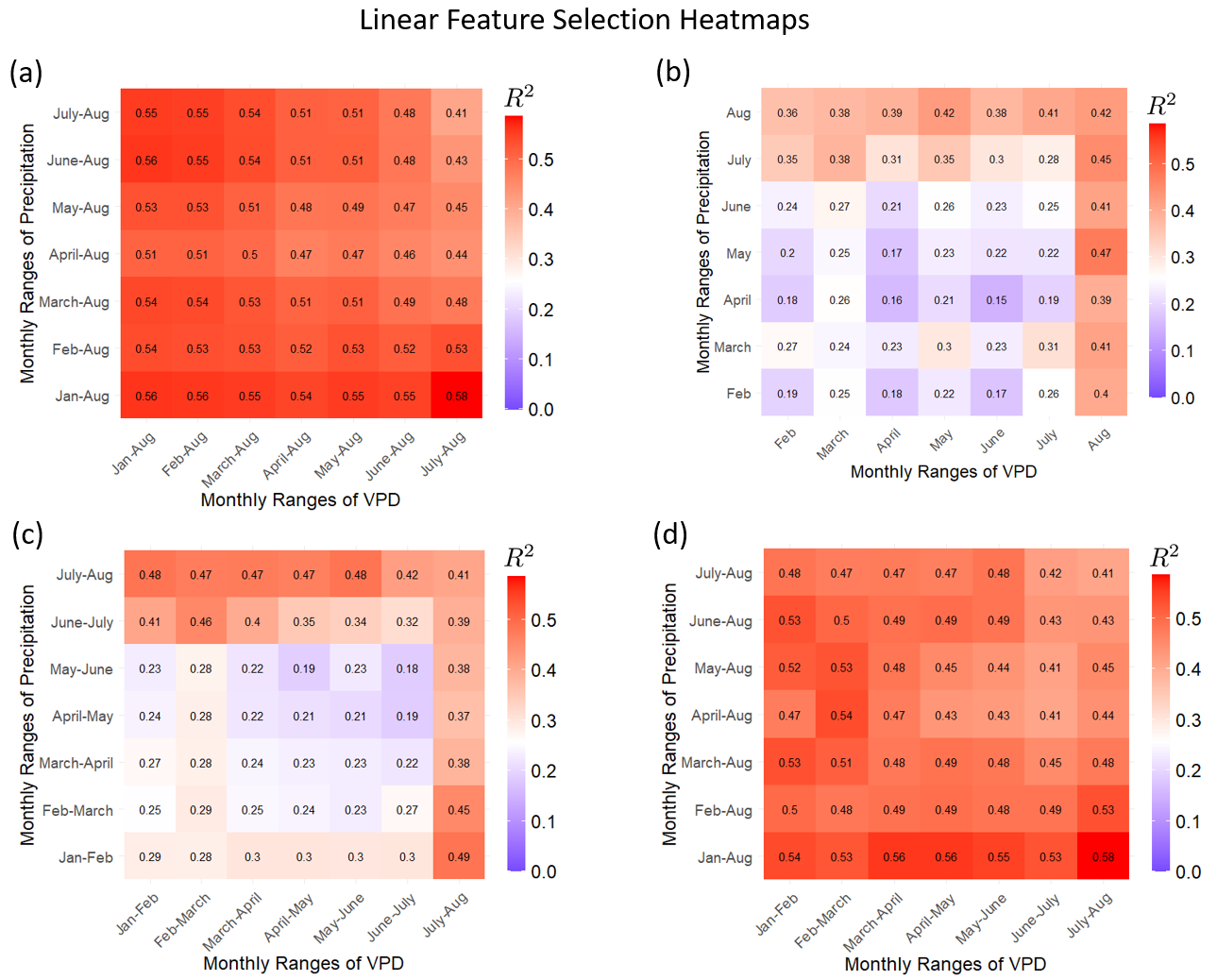}
    \caption{Overall $R^2$ values for various covariate combinations using training data and the linear attribution model. The horizontal axis of each heatmap lists the VPD monthly ranges. The vertical axes are the monthly ranges of precipitation. (a) Combinations include August VPD and August precipitation information. (b) Covariates of the linear model are only single months. (c) Covariates are an average across two months. (d) VPD averages are across two months, and precipitation averages include data up to August.}
    \label{fig:r2_supplement}
    
\end{figure}

Using model~\eqref{eq-lm-attr} 
and the first $n$ years of data as the initial training set, we experiment with varying temporal ranges of VPD and precipitation to model NDVI. As all models considered herein have the same number of variables, the overall $R^2$ is used to evaluate which combination of VPD and precipitation time ranges fit NDVI the best:

\begin{equation}
\label{grand_r2}
\text{Overall } R^2  =  \frac{1}{k}\sum_{i=1}^{k}\left(1- \frac{\sum_{t=1}^{n}(y_{i}(t_j)-\hat{y}_{i}(t_j))^2}{\sum_{t=1}^{n}(y_{i}(t_j)-\bar{y}_{i}(t_j))^2}\right).
\end{equation}
In Figure \ref{fig:r2_supplement}(a), we show the $R^2$ values from linear models using VPD and precipitation data from earlier months up to August. The combination of July–August Max VPD average and January-August average precipitation had the best fit, with an overall $R^2$ of 0.58. Figures \ref{fig:r2_supplement}(b)-(d) show the overall $R^2$ values for various other combinations of monthly ranges.

\textbf{Generalized Parallel Partial Gaussian Process.} An alternative approach to selecting the month ranges of precipitation and VPD is to compare the training marginal likelihoods across different combinations. We evaluate multiple G-PPGP models based on the structure in model~\eqref{general_ppgp}, where the covariates are defined by different monthly windows, whereby each combination yields a distinct covariance matrix. We then compute the marginal likelihood \citep{Gu2018robustness}, maximizing it with respect to the range parameter $\bm \gamma$ and the nugget parameter $\eta$. Due to the 
high computational cost, the analysis was performed based on a random subsample of 500 spatial locations. The top combinations based on the maximized marginal likelihood are shown in Table~\ref{lik_month_range}. The best-fitting G-PPGP model used average precipitation from January to August and average VPD from July to August. This result is in line with the $R^2$-based analysis.
\begin{table}[H]
\centering 
\caption{List of the top 5 covariate monthly range combinations ranked by maximized marginal likelihoods. The first 10 training years were used for likelihood estimation.}
\label{lik_month_range} 
\begin{tabular}{lll}
\toprule 
\textbf{Rank} & \textbf{Precipitation} & \textbf{VPD} \\
\midrule 
1 & January - August     & July - August \\ 
2 & January - August     & January - August \\ 
3 & January - August     & May - August \\ 
4 & January - August     & June - August \\ 
5 & January - August     & February - August \\ 

\bottomrule
\end{tabular}
\end{table}

\subsection[\appendixname~\thesubsection]{Model Specification and Hyperparameter Estimation for Attribution Models}\label{sec:hyperparam}

    \textbf{Generalized Parallel Partial Gaussian Process.} To estimate the range and nugget parameters of method~\eqref{general_ppgp}, we implement cross-validation. Suppose we use $n$ years of training data to forecast the August NDVI at year $t_{n+1}$. To further reduce computational cost, we first select a subset of locations, for which the indices are represented by a set $S = \{i_{1},i_{2},\ldots,i_{s}\}\subset \{1,\ldots,k\}$ with sample size $s<k$, i.e., $|S| = s$. Second, we separate the last $\nu$ years from the training set to use as a validation set. This validation set, with the length of $\nu<n$, is then used to calculate the RMSE, which is minimized based on the $n-\nu$ years of training data. The estimators are given as follows:
\begin{align*}
    (\hat{\bm \gamma}, \hat{\eta}) = \mbox{argmin}_{\gamma_{1},\gamma_{2},\eta>0}\left\{ \frac{1}{s\nu}\sum_{i\in S}\sum_{j = n-\nu+1}^{n}(y_{i}(t_j) - \hat{y}_{i}(t_j))^{2}  \right\},
\end{align*}
where the predictive mean $\hat{y}_{i}(t_j)$ is obtained from~\eqref{eq-gp-pred}, based on $n-\nu$ years of training history. We then use these estimators to project the August NDVI of each location at year $n+1$. The results in Table~\ref{tab:attribution_results}, \ref{tab:gross_ndvi_residual_table}, and \ref{tab:best_model_results} are based on $s = 500$ and $\nu=2$.

\textbf{Deep Neural Network (DNN).} Similar to model~\eqref{eq-lm-attr}, 
we train a separate DNN for each location.
For each location's independent neural network,  the input of the initial layer of the DNN 
is defined as $\mathbf{x}_{i}(t_j) = (x_{\text{precip},i}(t_j), x_{\text{vpd},i}(t_j))^{T}$, which is a vector of precipitation and VPD for the year $t_j$ at location $i$. 
For each layer $l$, the DNN contains $d_l$  neurons, where the output of these neurons is denoted as 
$\mathbf{h}_{l+1,i}(t_j)$, which takes the output of the previous layer's neurons, $\mathbf{h}_{l,i}(t_j)$, as input. For layer $l=0,...,L-1$ and location $i=1,...,k$:  
\begin{equation}
\mathbf{h}_{l+1, i}(t_j) = g \Bigl( \mathbf{W}_{l+1,i} \mathbf{h}_{l,i}(t_j) + \mathbf{b}_{l+1,i}\Bigr),
\end{equation}
where $\mathbf W_{l+1,i}$ is a $d_{l+1}\times d_l$ matrix of weights, and  $\mathbf b_{l+1,i}$ is a $d_{l+1}$ vector of bias parameters. Furthermore, $g$ represents an element-wise activation function for the hidden layer, such as rectified linear unit (ReLU), hyperbolic tangent function, and sigmoid function. In this analysis, we use the ReLU function. 
Lastly, the model output of the last layer is denoted as  
$y_{m,i}(t_j)=\mathbf W_{m,i}\mathbf h_{L,i}(t_j)+b_{L+1,i}$,  
where $\mathbf{W}_{m,i} \in \mathbb{R}^{1 \times d_{L}}$ is a row vector for the output layer, $b_{L+1,i}$ is the scalar-valued bias parameter for the output, and the subscript $m$ denotes "model" output.
The details about the number of neurons, layers, and the loss function to train the attribution DNN is in 
Table \ref{dnn_param}.

\begin{table}[H]
\centering 
\caption{Parameters for Deep Neural Network in attribution model. }
\label{dnn_param} 
\begin{tabular}{lll}
\toprule 
\textbf{Parameters} & \textbf{Values} &  \\
\midrule 
Dimension of the input & 2\\ 
The output dimension of the first fully connected layer & 64 \\ 
The output dimension of the second fully connected layer & 64  \\ 
The output dimension of the third fully connected layer & 1  \\ 
Batch size  & 1 \\
Learning rate & 0.001 \\
Epochs                        & 50     \\
\bottomrule 
\end{tabular}
\end{table}

\textbf{Fourier Neural Operator (FNO).}
For the attribution FNO, the input tensor  $ \mathbf{X}(t_j) \in \mathbb{R}^{k_1 \times k_2 \times 2} $ consists of two components representing the spatial maps of precipitation and VPD at time $t_j$, where $k_{1} = 100$ and $k_{2} = 140$. The model output matrix $ \mathbf{Y}_{m}(t_j) $ is the NDVI output that models the true NDVI $\mathbf Y(t_j) \in \mathbb{R}^{k_1 \times k_2}$. 
To train the FNO model, we first lift the input tensor $\mathbf{X}(t_j)$ to a higher-dimensional feature space by a linear transformation applied along the last dimension of the tensor. This is achieved by applying mode-3 tensor-matrix multiplication~\citep{kolda2009tensor} with a weight matrix $ \mathbf W_0 \in \mathbb{R}^{2 \times d_v} $, resulting in the transformed input $ \mathbf U_0 = \mathbf{X}(t_j) \times_3 \mathbf W_0 $, where $ d_v \geq 2 $. 
The transformed input $\mathbf U_0$ is then processed through $L$ successive Fourier layers. In each layer, FNO learns a mapping from $\mathbf U_l \in \mathbb{R}^{k_1 \times k_2 \times d_v}$ to $\mathbf U_{l+1} \in \mathbb{R}^{k_1 \times k_2 \times d_v}$ for $l = 1, \ldots, L$ : 
\begin{equation}\label{eq:fno_iteration}
\begin{aligned}
\mathbf U_{l+1} &= g \left( \mathbf U_l \times_3 \mathbf{W}_l + f_{v}(\mathbf U_l) + \mathbf{B}_l \right),
\end{aligned}
\end{equation}
where $g$ is the ReLU activation function acting on each element of the input comprised of three components: a point-wise linear transformation with weights $\mathbf{W}_l \in \mathbb{R}^{d_v \times d_v}$, a nonlocal convolution operator $f_{v}$ acting on the entire tensor $\mathbf U_l$, and a bias term $\mathbf{B}_l \in \mathbb{R}^{k_1 \times k_2 \times d_v}$. The convolution operator is parameterized by the 2D fast Fourier transform  \citep{cooley1965algorithm,duhamel1990fast}.  Lastly, the transformed representation $\mathbf U_L$ is projected to the output space through a linear transformation $\mathbf W_{L+1} \in \mathbb{R}^{d_v \times 1}$: 
$ \mathbf{Y}_m(t_j) = \mathbf U_L  \times_3 \mathbf W_{L+1}$. Table \ref{fno_param-a} shows the parameters for the attribution FNO.  

\begin{table}[H]
\centering 
\caption{Parameters for the attribution specific FNO NDVI model.}
\label{fno_param-a} 
\begin{tabular}{lll}
\toprule 
\textbf{Parameters} & \textbf{Values}  \\
\midrule 
The dimension, $d_v$, of the representation Fourier Layer $v(i)$ & 16  \\
The max Fourier modes used in the Fourier layer & 50  \\
The resolution of the discretization in the first dimension & 140 \\
The resolution of the discretization in the second dimension & 100\\
Batch size& 1 \\
Learning rate & 0.001\\
Epochs & 100 \\
\bottomrule 
\end{tabular}
\end{table}

\subsection[\appendixname~\thesubsection]{Model Specification and Hyperparameter Estimation for Forecast Models}\label{sec:forecast_hyperparam}

\textbf{Parallel Partial Gaussian Process (PPGP).} For the PPGP method in Section~\ref{subsec:one-year-ahead_cov}, given $n$ observations of training data and hyperparameters $\rho_{q}$ and $\eta_{q}$, a non-central Student's t-distribution with $n-1$ degree of freedom can be utilized to obtain a predictive mean and deviance at time $t_{*}$:
\begin{align}\label{eq-gp-pred-x}
    (x_{q,i}(t_{*})\mid \mathbf x_{q,i}(t_{1:n}), \rho_{q}, \eta_{q})\sim \mathcal{T}(\hat{x}_{q,i}(t_{*}), \hat{\sigma}_{q,i}^{2}K_{q}^{*}, n-1),
\end{align}
which parallels Equation \ref{eq-gp-pred} of the manuscript. 
Here, $q$ indicating either "precip" or "vpd". In comparison to Equation \ref{eq-gp-pred}, instead of $\hat{y}_{i}(\mathbf x_i(t_{*}))$, we have $\mathbf x_{q,i}(t_{1:n})$, which is an $n$-dimensional vector $(x_{q,i}(t_1), x_{q,i}(t_2),\ldots,x_{q,i}(t_n))^{T}$. Additionally, instead of $\hat{\mu}_{i}$, we have $\hat{\mu}_{q,i} = \left(\mathbf{1}_{n}^{T}\Tilde{\mathbf K}_{q}^{-1}\mathbf{1}_{n}\right)^{-1}\mathbf{1}_{n}^{T}\Tilde{\mathbf K}_{q}^{-1}\mathbf{x}_{q,i}(t_{1:n})$ where $K_{q}^{*}$ is computed in a similar manner as~\eqref{eq:pred_cov} but with $\mathbf k_{q}(t_{*}) = [\mbox{Cor}(x_{q, i}(t_1),  x_{q, i}(t_{*})),\mbox{Cor}(x_{q, i}(t_2),  x_{q, i}(t_{*})),\ldots,\mbox{Cor}(x_{q, i}(t_n),  x_{q, i}(t_{*}))]^{T}$ based on covariate-specific correlation functions defined in Section \ref{subsec:one-year-ahead_cov}. Similar to G-PPGP, a validation set is used to estimate the parameters. With a given subsample $S_{q} = \{i_{q,1},\ldots, i_{q,s}\}\subset \{1,\ldots,k\}$ with size $s_{q}$ and the length of the validation set $\nu_{q}<n$, the estimator follows

\begin{align*}
    (\hat{\rho}_{q}, \hat{\eta}_{q}) = \mbox{argmin}_{\rho_{1}\in(-1,1), \rho_{2},\eta_{1},\eta_{2}>0}\left\{ \frac{1}{s_{q}\nu_{q}}\sum_{i\in S_{q}}\sum_{j = n-\nu_{q}+1}^{n}(x_{q,i}(t_j) - \hat{x}_{q,i}(t_j))^{2}  \right\},
\end{align*}
where the mean forecast $\hat{x}_{q,i}(t_j)$ is calculated in a manner similar to that described in~\eqref{eq-gp-pred-x}, with $n-\nu$ years of training. 
We used $s_{1}=s_{2}=10$, $\nu_{1}=2$, and $\nu_{2}=3$ to obtain the results in Table \ref{tab:precip_forecast}, Table \ref{tab:vpd_forecast}, 
and Figure \ref{fig:vpd_results}. 

\textbf{Fourier Neural Operator (FNO). }
We use FNO to predict August NDVI, VPD, and Precipitation one year ahead. We use a similar architecture to the FNO model in Section~\ref{subsec:attribution}, except that the input matrix is $\mathbf{Y}(t_{j-1}) \in \mathbb{R}^{k_1 \times k_2}$, where $\mathbf{Y}(t_{j-1})$ represents the spatial map of the variable to be predicted in the previous year for $j=2,\ldots,n$. Hyperparameters and configuration of the NDVI and precipitation forecast using FNO is listed in Table \ref{fno_param-f}. The hyperparameters for the VPD FNO forecast model are listed in Table \ref{fno_param-f-vpd}.

\begin{table}[H]
\centering 
\caption{Parameters for the Fourier Neural Operator model for forecasting NDVI and precipitation one year ahead, where input is previously known NDVI or precipitation from prior years, respectively. }
\label{fno_param-f} 
\begin{tabular}{lll}
\toprule 
\textbf{Parameters} & \textbf{Values} \\
\midrule 
The dimension, $d_v$, of the representation Fourier Layer $v(i)$& 16   \\
The max Fourier modes used in the Fourier layer & 50\\
The resolution of the discretization in the first dimension & 140\\
The resolution of the discretization in the second dimension & 100 \\
Batch size& 1\\
Learning rate & 0.001\\
Epochs& 100\\
\bottomrule 
\end{tabular}
\end{table}

\begin{table}[H]
\centering 
\caption{Parameters for the Fourier Neural Operator model of VPD one year ahead forecasting, where input is previously known VPD from prior years. }
\label{fno_param-f-vpd} 
\begin{tabular}{lll}
\toprule 
\textbf{Parameters} & \textbf{Values}  \\
\midrule 
 The dimension, $d_v$, of the representation Fourier Layer $v(i)$ & 16 \\
The max Fourier modes used in the Fourier layer & 16 \\
The resolution of the discretization in the first dimension & 140 \\
The resolution of the discretization in the second dimension  & 100 \\
Batch size& 1 \\
Learning rate & 0.001\\
Epochs & 100\\
\bottomrule 
\end{tabular}
\end{table}

\textbf{Recurrent Neural Network (RNN).} At each location $i$, the RNN is a sequence of hidden states, $\mathbf h_{i}(t_j)$ for $j = 1, 2, \ldots, n$. At time step $t_j$, the current hidden state $\mathbf{h}_i(t_j)$ takes vector $\mathbf{x}_{i}(t_{j-1})$ and the previous time step's hidden state $\mathbf{h}_{i}(t_{j-1})$ as inputs: 
\begin{equation}
\label{eq:RNN}
\begin{aligned}
\mathbf{h}_i(t_j) = g(\mathbf{W}_{h,i} \mathbf{h}_{i}(t_{j-1}) + \mathbf{W}_{x,i} \mathbf{x}_i(t_{j-1}) + \mathbf{b}_{h,i}(t_j)),
\end{aligned}
\end{equation}
where $\mathbf W_{h,i}\in \mathbb{R}^{d_{h}\times d_{h} }$ and $\mathbf W_{x,i}\in \mathbb{R}^{2 \times d_{h}}$ represent weights, $\mathbf{x}_i(t_{j-1})$ is the input vector comprised of precipitation and VPD at year $t_{j-1}$, and $\mathbf b_{h,i}(t_j)\in\mathbb R^{d_{h}}$ represents the bias. Here, $d_{h}$ is the dimension of all hidden states. The initial input, $\mathbf{h}_i(1)$, is defined as a zero vector with length $d_{h}$. We use the  element-wise ReLU function for the activation function $g$. The model output is denoted as $y_{m,i}(t_j) = \mathbf{W}_{y,i} \mathbf{h}_i(t_j) + b_{m,i}$, where $\mathbf{W}_{y,i} \in \mathbb{R}^{1 \times d_{h}}$ is a row vector for the output layer, $b_{m,i}$ is the scalar-valued bias parameter for the output, and the subscript "m" denotes the "model" output. Hyperparameter specifications for the RNN method are in Table \ref{rnn_param}.

\begin{table}[H]
\centering 
\caption{Parameters for Recurrent Neural Network one year ahead forecasting. }
\label{rnn_param} 
\begin{tabular}{lll}
\toprule 
\textbf{Parameters} & \textbf{Values} \\
\midrule 
 Dimension of the input & 2\\ 
Dimension of the RNN's hidden layers & 64 \\ 
Output dimension of the first fully connected layer & 64\\ 
Output dimension of the second fully connected layer & 64 \\ 
Output dimension of the third fully connected layer & 1 \\ 
Batch size & 1\\
Learning rate & 0.001\\
Epochs & 50\\
\bottomrule 
\end{tabular}
\end{table}

\section{Long-Term Forecasts of Climate Attributes Results}\label{cov_forecast_supp}

\begin{figure}[H]
    \centering
    \includegraphics[width=1\textwidth]{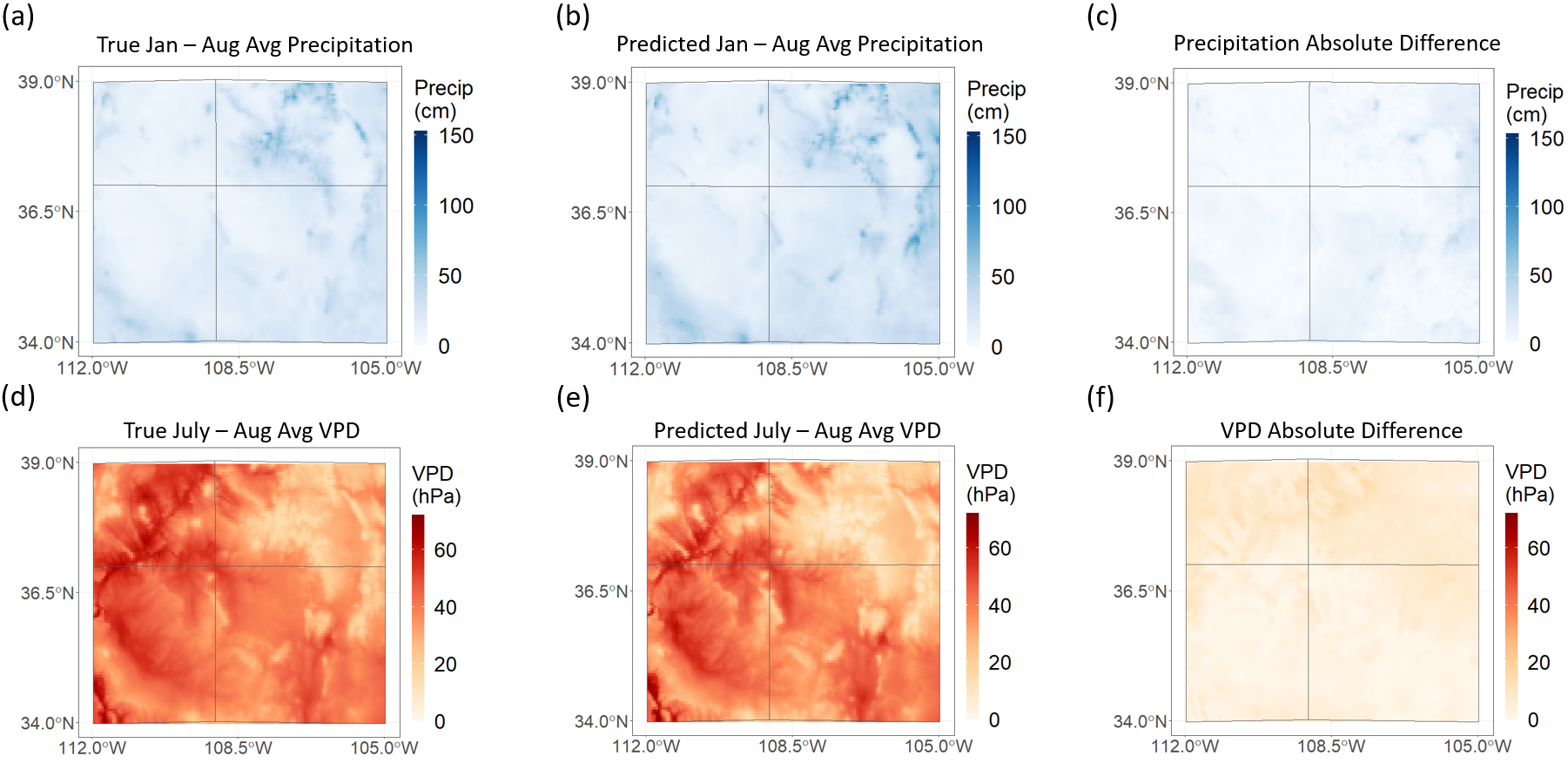}
    \caption{(a) True average precipitation between January and August  of the Four Corners region in the year 2020. (b) The one-year-ahead predicted January-August average precipitation using the PPGP method in  the year 2020. (c) Absolute value of the true precipitation minus the predicted values for the year 2020. (d) True average max VPD during the months of July-August in the year 2020. (e) The PPGP one-year-ahead prediction of max VPD average between July-August for the year 2020. (f)  The absolute value of the true VPD minus the predicted values for the year 2020.}    
   \label{fig:cov_results}  
\end{figure}

In Figure \ref{fig:cov_results}, 
we plot the true one-year-ahead predicted January-August precipitation and July-August VPD in 2020, their forecasts by PPGP, and the absolute difference between the truth and the prediction, $|y_i(t_j) - \hat{y}_i(t_j)|$ for $i= 1, ..., k$ with $k$ being the number of grids, and $j=n+n^*$ (corresponding to the year 2020). 
The predicted values of precipitation and VPD are closely aligned with the actual 2020 values, as shown in  Figure \ref{fig:cov_results}(c) and Figure \ref{fig:cov_results}(f), respectively. 

Table \ref{tab:precip_forecast} shows the results of forecasting precipitation one year ahead using various methods. The PPGP method performs the best with the smallest RMSE. The $P_{95\%}$ coverage is larger than the LM method, and the $L_{95\%}$ is significantly smaller, indicating precise uncertainty quantification when using the PPGP method. LM and FNO do not perform as well due to the inflexibility of linear models and the lack of data for a deep learning approach. Using the previous year's January-August precipitation does not perform well due to the negative correlation between consecutive years. 
\begin{table}[H]
		\caption{Precipitation forecast accuracy and assessment for PPGP, linear model, FNO, historical location average, and using the previous year as prediction. The RMSE as well as the coverage and average length of the 95\% CI are reported for the PPGP and LM methods. 
        }	
		\label{tab:precip_forecast}
		\centering
		\begin{tabular}{lccccc}
			\hline
	Precip. Forecast & PPGP & LM & FNO  & Location Mean & Previous Yr \\
			\hline  
			RMSE  &  \textbf{8.24} & 10.4 & 12.1 & 9.21 & 14.2\\
            $P_{95\%}$ & \textbf{0.915} &0.911 & --- & --- & ---\\
            $L_{95\%}$ & \textbf{28.1} & 38.1 & --- & --- & ---\\
            
           
			\hline  
		\end{tabular}
\end{table}

 Table \ref{tab:vpd_forecast} contains the performance metrics of VPD forecasting when predicting the last 8 years of data. FNO does not perform well, as this is a situation with very little data. PPGP has a smaller RMSE than the historic location mean and has a good $P_{95\%}$ coverage. However, due to VPD's overall trend changes between the training set and test set, using the previous year's July-August VPD at each location as the prediction performs slightly better when predicting the last 8 years of data. 

\begin{table}[H]
		\caption{VPD forecast accuracy and assessment for PPGP, FNO, location mean, and previous year methods. Average RMSE, coverage, and average length of interval are reported.
        }	
		\label{tab:vpd_forecast}
		\centering
		\begin{tabular}{lccccc}
			\hline
	VPD Forecast & PPGP & 
    FNO & Location Mean  & Previous Yr\\ 
			\hline  
			RMSE  & \textbf{3.46} & 
            3.66 &4.02 & 3.04\\
            $P_{95\%}$ & 0.960 
            & --- & --- & ---\\
            $L_{95\%}$ & 18.0 
            & --- & --- & ---\\
           
			\hline  
		\end{tabular}
\end{table}

\begin{figure}[H]
    \centering
    \includegraphics[width=0.68\textwidth]{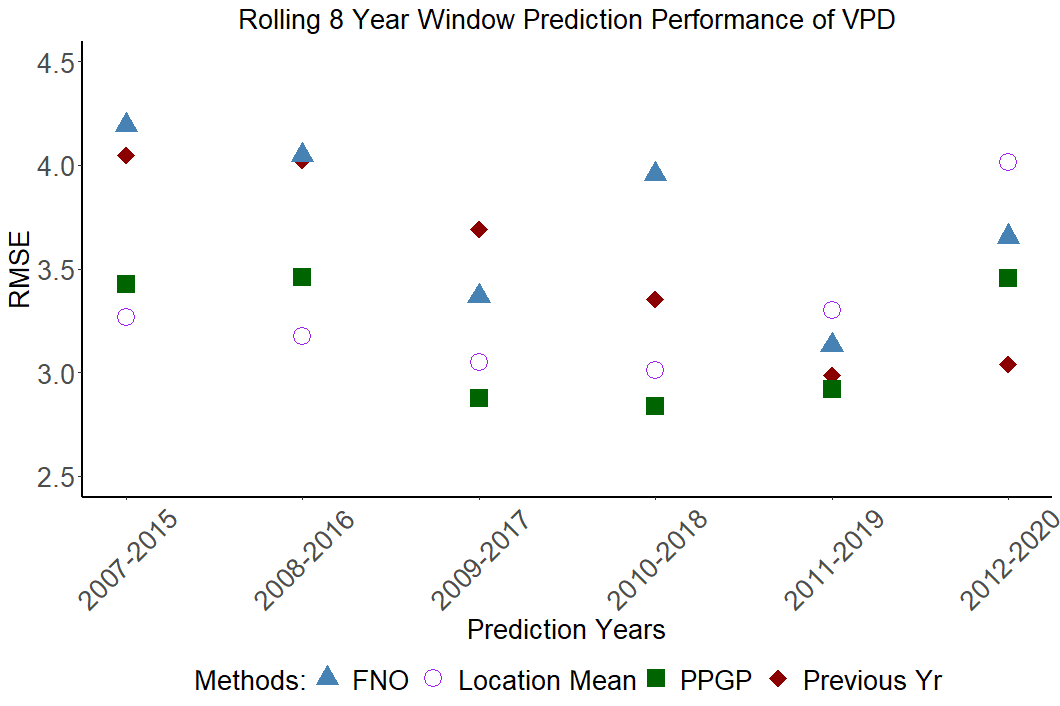}
    \caption{
    RMSEs for one-year-ahead forecasts of VPD when predicting an 8-year rolling window using FNO, historic location mean, PPGP, and previous year (prediction is the previous year's July-August VPD average at each location) methods. The average RMSE for each of the methods is 3.79, 3.30, 3.16, and 3.52, respectively.
    }
   \label{fig:vpd_results}  
\end{figure}

To further explore the prediction of VPD, Figure \ref{fig:vpd_results} displays the RMSE for 
a sequence of 8-year rolling window predictions starting from the year 2007. As shown, PPGP consistently performs well compared to the other methods. When using the historical location VPD mean, as the prediction of VPD, the prediction becomes worse in later years because the location mean is not flexible to recent drastic changes in VPD. The previous year method (where the VPD prediction is the previous year's July-August VPD average at each location) performs poorly during the earlier windows, but performs well in the more recent windows. 
The FNO method performs poorly in many of the prediction windows, as there is very little data for a deep learning method to work well. Even though the PPGP method is not the best performing for the latest window, 
the PPGP method performs better on average compared to all other methods for different rolling windows in Figure \ref{fig:vpd_results}. 

\subsection{Additional NDVI Attribution and One-Year-Ahead Results} \label{sec:add_att_models}

\begin{figure}[H]
    \centering
    \includegraphics[width=1\textwidth]{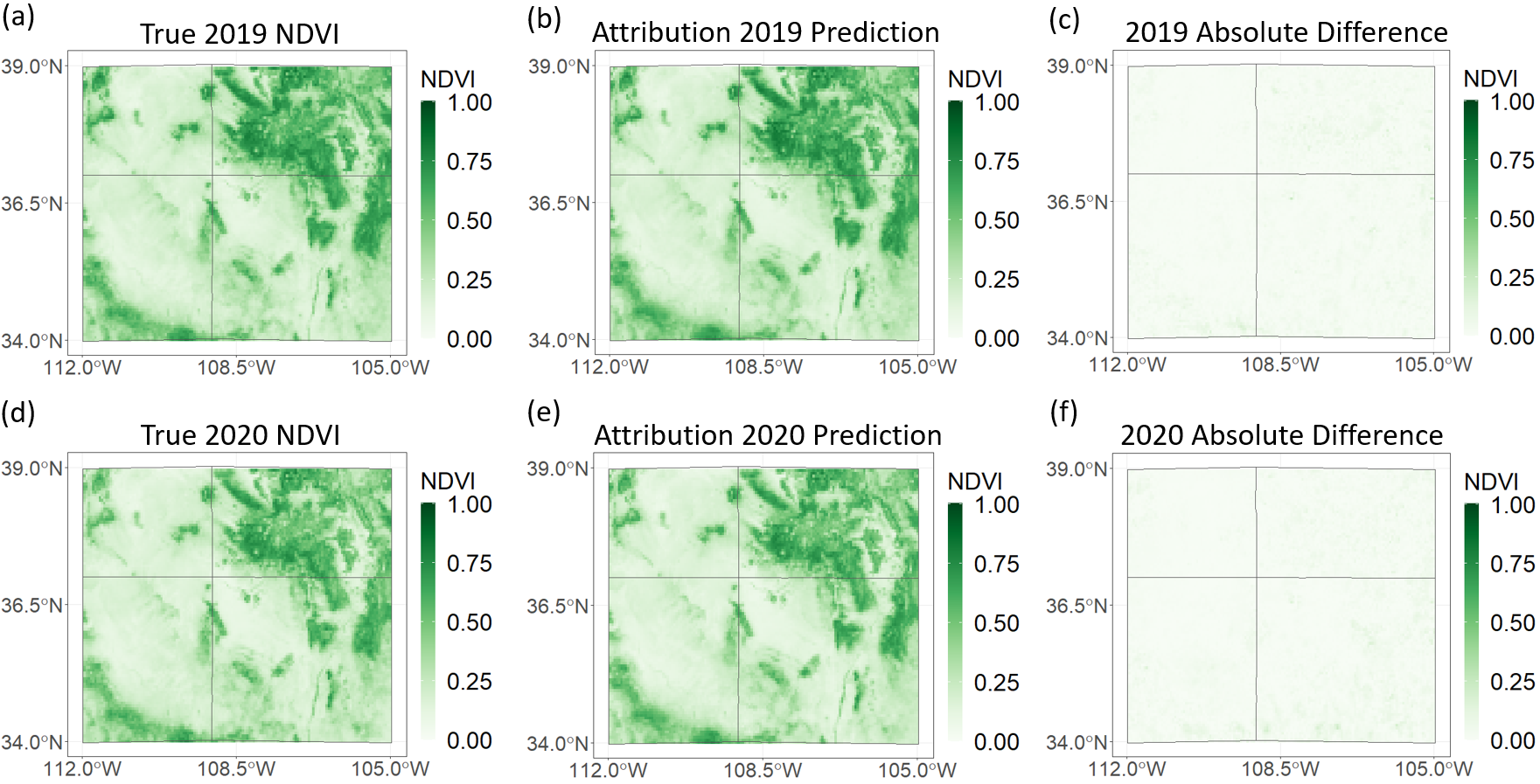}
    \caption{(a) True 2019 August NDVI. (b) G-PPGP attribution prediction of 2019 August NDVI. (c) Absolute difference between truth and attribution prediction for the year 2019. (d) True 2020 August NDVI. (e) G-PPGP attribution prediction of 2020 August NDVI. (f) Absolute difference between truth and attribution prediction for the year 2020. }
    \label{fig:2019_2020_attribution}
\end{figure}

Figure \ref{fig:2019_2020_attribution} shows the true 2019 and 2020 NDVI values in the Four Corners region, the G-PPGP attribution predictions of NDVI for 2019 and 2020, and the absolute value of truth minus the attribution prediction; the prediction closely resembles the truth. 

\begin{figure}[H]
    \centering
    \includegraphics[width=1\textwidth]{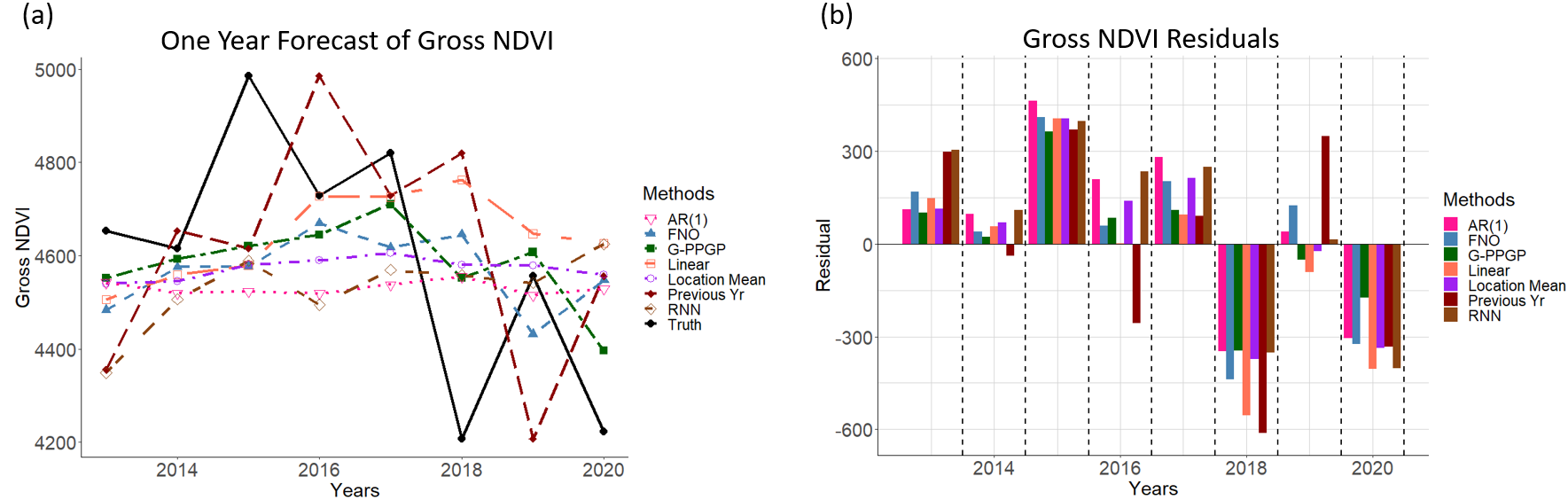}                 
    \caption{(a) The true gross NDVI, calculated as the sum of all August NDVI values over all grids for each prediction year, and the one-year-ahead forecasted gross NDVI using all  methods. (b) The residuals between true gross NDVI and forecasted gross NDVI for all forecasting methods.}
    \label{fig:full_gross_ndvi}
\end{figure}

Figure \ref{fig:full_gross_ndvi} presents the gross NDVI (sum of all August NDVI values across all grids for a given year) one-year-ahead prediction using methods discussed in Section~\ref{ndvi_forecast} of the manuscript. In general, G-PPGP has the most accurate predictions.

\section{Short-Term Forecasting Using Monthly Data} \label{sec:less_than_one_yr}

Once the covariates and NDVI are predicted one year into the future, more information can be added throughout the prediction year as it is observed. Specifically, to predict August NDVI in year $t_{n+1}$, we used up to $n$ years of data, but as year $t_{n+1}$ is observed, information about precipitation during this year can be included to obtain a more accurate prediction. Thus, for each training year $j=1,\cdots,n$, we consider simple linear regression models of January-August average precipitation on January, January-February, January-March, ..., and 
January-July average precipitation:

\begin{align}\label{eq-preci-sptial-model}
    x_{\text{precip},i}(t_j) = \beta_{0}(t_j,r) + \beta_{1}(t_j,r)x_{\text{precip},i}(t_j,r) + \epsilon_{i}(t_j,r),
\end{align}
where $i=1,\cdots,k$, and $x_{\text{precip},i}(t_j,r)$ refers to the average precipitation of the first $r$ months (prior to August) of the current year with $r=1,\cdots,7$. Additionally, $\epsilon_{i}(t_j,r)$ is the independent Gaussian white noise with variance $\sigma^{2}(t_j,r)$.  For a fixed year $t_j$, the model in equation~\eqref{eq-preci-sptial-model}, fits a linear model across different spatial locations. Once the linear model coefficients for the first $r$ months are fit for the training years, denoted as $(\hat\beta_{0}(t_j,r), \hat\beta_{1}(t_j,r))^{T}$ for $j=1,\cdots,n$, we use a summarized pair of coefficients for the test years:
\begin{equation}\label{eq-preci-sptial-model-coeff}
    \begin{aligned}
    \hat\beta_{0}^{\text{Train}}(r) & = \frac{1}{n}\sum_{j=1}^{n}\hat\beta_{0}(t_j,r),\\
    \hat\beta_{1}^{\text{Train}}(r) & = \frac{1}{n}\sum_{j=1}^{n}\hat\beta_{1}(t_j,r).    
    \end{aligned}
\end{equation}
The superscript "Train" in~\eqref{eq-preci-sptial-model-coeff} denotes that the coefficients are obtained from training years. For example, for a given future year, $j = n+1,\cdots,n+n^{*}$, the January-August average precipitation given the January-March average precipitation of that future year at the $i$-th location would be predicted by
\begin{align}\label{eq-preci-pred}
    \hat{x}_{\text{precip},i}^{(\text{S})}(t_{*}) = \hat\beta_{0}^{\text{Train}}(3) + \hat\beta_{1}^{\text{Train}}(3)x_{\text{precip},i}(t_{*},3),
\end{align}
where the superscript "S" in~\eqref{eq-preci-pred} refers to a predicted value based on the "spatial" regression model in~\eqref{eq-preci-sptial-model}, and $t_*$ is a prediction year in the test set. Furthermore, let  $\hat{x}_{1,i}^{(\text{P})}(t_{*})$ denote the predicted precipitation from PPGP described in~\eqref{eq-gp-pred-x} with the superscription "P" representing "PPGP". By employing both $\hat{x}_{\text{precip},i}^{(\text{P})}(t_{*})$ and $\hat{x}_{\text{precip},i}^{(\text{S})}(t_{*})$ as a weighted average, we have another prediction of precipitation:
\begin{align}\label{eq-preci-weighted}
    \hat{x}_{\text{precip},i}(t_{*}) = w\hat{x}_{\text{precip},i}^{(\text{P})}(t_{*}) + (1-w)\hat{x}_{\text{precip},i}^{(\text{S})}(t_{*}),\ w\in(0,1),
\end{align}
which incorporates both temporal dynamics and spatial trends of precipitation. The weight $w$ may be chosen under the training-period loss function which is defined as
\begin{align*}\label{eq-loss-train-period}
    L(w) = \sum_{i=1}^{k}\sum_{j=1}^{n}\left(x_{\text{precip},i}(t_j)-\hat{x}_{\text{precip},i}(t_{*})\right)^{2},
\end{align*}
where the optimal weight $\hat{w}$ is obtained by $\hat{w}  = \mbox{argmin}_{w\in(0,1)}L(w)$, with different choices of the ending month $r=1,\cdots,7$ within the current year.
We use the weighted prediction of precipitation, as described in~\eqref{eq-preci-weighted}, and the G-PPGP model to predict August NDVI in a time frame smaller than one-year-ahead. Table~\ref{tab:weighted-prediction} presents the weights and the performance of prediction for precipitation and NDVI when different month range $r$ is used. As we include more information of the current year by increasing $r$, the weight decreases, and the general predictive performance increases for both precipitation and NDVI. The forecasts become more accurate by including more information within the prediction year, but it suggests that the prediction gain might not be large until most of the truth is observed (knowing the January-July average precipitation of that year). 
\begin{table}[H]
		\caption{Performance of the weighted forecast of January-August precipitation and August NDVI within a prediction window of less than one year. The "RMSE (PPGP)" column reports the root mean square errors of PPGP  precipitation forecasting, while the "RMSE (G-PPGP)" column reports the root mean square errors of G-PPGP NDVI forecasting, where the weighted precipitation estimates are used as inputs. The $\hat{w}$ column shows the optimal weight, and $r$ indicates which month of the current year we include information up to. 
  }	
		\label{tab:weighted-prediction}
		\centering
		\begin{tabular}{cccc}
			\hline
	    $r$ & $\hat{w}$ & RMSE (PPGP) & RMSE (G-PPGP)\\
			\hline  
   1 & 0.62 & 8.16 & 0.0348\\
   2 & 0.58 & 7.42 & 0.0352\\
   3 & 0.50 & 7.55 & 0.0356\\
   4 & 0.41 & 7.23 & 0.0353\\
   5 & 0.37 & 6.05 & 0.0342\\
   6 & 0.30 & 5.45 & 0.0340\\
   7 & 0.11 & 3.37 & 0.0322\\
			\hline
		\end{tabular}
	\end{table}

\section{Additional Model Comparison}\label{additional_appendix_models}

Table \ref{tab:prev_vdp_prev_precip_results} shows the results when only historical (years $t_1$ to $t_{j-1}$) is used to predict NDVI in year $t_j$, and no forecasted covariates are included as inputs. The RMSE values for both G-PPGP and LM are larger than results in Table \ref{tab:best_model_results}, where forecasted VPD and precipitation are also included as inputs.

\cite{funk2006intra} forecasted NDVI one to four months ahead in semi-arid Africa and found that precipitation, relative humidity, and NDVI were important for one-month forecasts; Table \ref{tab:attribution_prevNDVI_results}  explores the inclusion of NDVI as a covariate when forecasting one year ahead. We found that adding peak NDVI from prior years does not improve prediction accuracy, as the information from peak NDVI in previous years is not as informative as NDVI measured one month prior to the prediction date. Nevertheless, the proposed G-PPGP models can incorporate additional covariates and potentially be applied to other regions of the world.


\begin{table}[H]
		\caption{G-PPGP and linear models where the inputs are only historic VPD and historic precipitation.  }	
		\label{tab:prev_vdp_prev_precip_results}
		\centering

        \begin{tabular}{lcccc}
            \hline 
           
	Method & Input & RMSE & P(95\%) & L(95\%)  \\
			\hline
	       G-PPGP  & Historic VPD/Precip & \textbf{0.0370} & 0.932  & \textbf{0.137}\\
	       LM  &  Historic VPD/Precip &0.0425 &\textbf{0.942} &  0.158\\
			\hline
		\end{tabular}
	\end{table}

\begin{table}[H]
		\caption{G-PPGP and linear attribution models where VPD is known, precipitation is known, and the previous August NDVI values are additional inputs. }	
		\label{tab:attribution_prevNDVI_results}
		\centering

        \begin{tabular}{lcccc}
            \hline 
           
	Method & Input & RMSE & P(95\%) & L(95\%)  \\
			\hline
	       G-PPGP  & Known VPD/Precip and Last Aug NDVI & \textbf{0.0294} & \textbf{0.952}  & \textbf{0.113}\\
	       LM  &  Known VPD/Precip and Last Aug NDVI & 0.0327 & 0.944 &  0.127\\
			\hline
		\end{tabular}
	\end{table}



\begin{adjustwidth}{-\extralength}{0cm}

\reftitle{References}


\bibliography{References}


%


\PublishersNote{}
\end{adjustwidth}
\end{document}